\documentclass[12pt,a4paper]{article}

\usepackage[truedimen,margin=30mm]{geometry}
\usepackage[]{graphicx}
\usepackage{authblk}   
\usepackage{xurl}
\usepackage{setspace}
\usepackage{placeins}
\usepackage{microtype}

\usepackage[]{natbib}

\usepackage{amsmath}
\usepackage{appendix}
\usepackage{dsfont}
\usepackage{subcaption}
\usepackage{multirow}
\usepackage[hidelinks,hyperfootnotes=false]{hyperref}
\usepackage{orcidlink}
\usepackage{float}

\newcommand{\relmiddle}[1]{\mathrel{}\middle#1\mathrel{}}

\usepackage{color}

\setcitestyle{open={[},close={]}}

\makeatletter
\def\mojiparline#1{
    \newcounter{mpl}
    \setcounter{mpl}{#1}
    \@tempdima=\linewidth
    \advance\@tempdima by-\value{mpl}zw
    \addtocounter{mpl}{-1}
    \divide\@tempdima by \value{mpl}
    \advance\kanjiskip by\@tempdima
    \advance\parindent by\@tempdima
}
\makeatother
\def\linesparpage#1{
    \baselineskip=\textheight
    \divide\baselineskip by #1
}

\allowdisplaybreaks[4]

\title{ {\bf Alternative prior specifications and extension to hybrid-controlled trials in Bayesian designs using external data } }

\author[1,2]{\orcidlink{0009-0001-1986-5325}\,Wataru Murasaki \footnote{Corresponding author: Wataru Murasaki, Department of Biostatistics, Tsukuba Clinical Research \& Development Organization, University of Tsukuba, 2-1-1 Amakubo, Ibaraki 305-8576, Japan. E-mail: w.murasaki.vios@gmail.com}}

\author[3]{\orcidlink{0000-0002-9630-5882}\,Tomohiro Ohigashi}

\author[4]{\orcidlink{0000-0003-1725-4579}\,Ryota Ishii}

\author[4]{\orcidlink{0000-0002-3093-0386}\,Kazushi Maruo}

\author[4]{\orcidlink{0000-0002-5973-9163}\,Masahiko Gosho}

\affil[1]{Graduate School of Comprehensive Human Sciences, University of Tsukuba, Ibaraki, Japan}

\affil[2]{Department of Biostatistics, Tsukuba Clinical Research \& Development
    Organization, University of Tsukuba, Ibaraki, Japan}

\affil[3]{Department of Information and Computer Technology, Faculty of Engineering, Tokyo University of Science, Tokyo, Japan}

\affil[4]{Department of Biostatistics, Institute of Medicine, University of Tsukuba, Ibaraki, Japan}

\date{}

\begin{document}
\linesparpage{25}
\allowdisplaybreaks[4]
\begin{singlespace}
	\maketitle
\end{singlespace}

\vspace{-1cm}
\begin{center}
	{\large\bf Abstract}
\end{center}
Bayesian methods incorporating external data are attractive for clinical trials with limited patient availability, but controlling the type I error rate remains challenging.
Psioda and Ibrahim (2019) proposed a simulation-based framework using external data, which evaluates the Bayesian type I error and power as rejection probabilities averaged over sampling priors.
However, when the external data show a large treatment effect, the null sampling prior can degenerate and the Bayesian type I error rate approaches the frequentist rate.
Moreover, the robust mixture prior used as the analysis prior has a second component as informative as the external data, potentially biasing estimates, and the framework does not accommodate hybrid-controlled trials.
We propose an alternative sampling prior specification, assess estimation bias under the analysis prior, and extend the framework to hybrid-controlled trials.
In simulations, the existing specification required sample sizes comparable to or larger than those of frequentist designs, whereas the proposed specification reduced the required sample size but allowed the frequentist type I error rate to exceed the nominal level.
A weakly informative second component also reduced estimation bias under prior-data conflict.
A pediatric cutaneous lupus erythematosus case study illustrates the method.


\section{Introduction}\label{sec1}
Determining an appropriate sample size for a clinical trial based on hypothesis testing is crucial to ensure sufficient statistical power while maintaining the type I error rate \citep{ICHE9, andrade-SampleSizeIts-2020}.
However, in trials targeting rare diseases and pediatric populations, recruiting a sufficient number of patients to meet the sample size requirements is often challenging.

To address this issue, various methods incorporate external (historical) data---such as existing clinical trial or real-world data---into current (new) trial data analysis.
The Bayesian method is an attractive approach for incorporating external data because it naturally integrates such data through a prior distribution \citep{spiegelhalter-BayesianApproachesClinical-2004}.
Regulatory agencies have increasingly recognized the potential benefits of incorporating external data into clinical trials.
The Food and Drug Administration (FDA) has released guidelines on Bayesian statistics in medical device clinical trials \citep{FDA-in-MedicalDevice-2010}, on natural history studies for rare diseases \citep{RareDiseasesNatural}, and in draft form, on externally controlled trials \citep{ConsiderationsDesignConduct}.
The International Council for Harmonisation (ICH) has also issued the ICH E11A guideline, which describes the use of data from adults or other reference populations in pediatric drug development \citep{ICHGuidelineE11A2013}.
Recently, the FDA also published draft guidance on Bayesian methodology for drugs and biological products, which discusses borrowing external data through informative priors and the evaluation of operating characteristics \citep{FDA-in-DrugDevelopment-2026}.

When using the Bayesian approach to trial design, determining the required sample size for the current trial is essential.
Several criteria based on parameter inference performance have been proposed for Bayesian sample size determination.
For example, traditional approaches focus on the posterior performance of parameters, using metrics such as the Average Length Criterion (ALC) \citep{joseph-BayesianSampleSize-1997} and the Average Posterior Variance Criterion (APVC) \citep{gelfand-SimulationbasedApproachBayesian-2002}, which aim to ensure the precision of posterior estimates \citep{brutti-BayesianfrequentistSampleSize-2014}.
Sample size determination methods based on decision-theoretic approaches \citep{lindley-ChoiceSampleSize-1997} and Bayes factors \citep{weiss-BayesianSampleSize-1997} have also been proposed.
Alternatively, sample size can be determined by minimizing a weighted sum of the type I and type II error rates, which decrease as the sample size increases \citep{grieve-HowTestHypotheses+2015, pericchi-AdaptativeSignificanceLevels+2016}.
In addition, sample size determination methods based on the frequentist approach to hypothesis testing have been proposed and have gained attention, particularly in the regulatory context \citep{gelfand-SimulationbasedApproachBayesian-2002, chen-BayesianDesignNoninferiority-2011, FDA-in-DrugDevelopment-2026}.
However, several studies \citep{kopp-schneider-PowerGainsUsing-2020, best-ClassicalTypeError-, lee-ElicitingDiscountParameter-2024} have noted that controlling the type I error rate at the same nominal level as used in frequentist designs becomes difficult when external data are incorporated into the analysis of the current trial data.

\citet{psioda-BayesianClinicalTrial-2019} proposed a potential solution to this issue with a simulation-based sample size determination method within a Bayesian framework using two priors, called the sampling prior and analysis prior, based on the work of \citet{gelfand-SimulationbasedApproachBayesian-2002} and \citet{chen-BayesianDesignNoninferiority-2011}.
The sampling prior is used to generate current trial data under the null and alternative hypotheses, and is referred to as the null sampling prior (NSP) and alternative sampling prior (ASP), respectively.
During the design phase of a clinical trial, data consistent with the null and alternative hypotheses are generated for each sampling prior.
An analysis prior is then used to calculate the posterior distribution, updated by the data generated from the NSP and ASP, and to estimate the treatment effect.
The Bayesian type I error rate and power are defined as the probabilities of rejecting the null hypothesis under the NSP and ASP, respectively.
The required sample size for their method is the minimum number of patients needed to achieve the desired Bayesian power while maintaining the Bayesian type I error rate at a nominal level.
The key idea of their framework is to relax strict frequentist type I error control, that is, the requirement that the type I error rate not exceed the nominal level at any parameter value in the null space.
Instead, the type I error rate is controlled on average over the NSP.
Under this criterion, the type I error rate may exceed the nominal level for some parameter values in the null space, but its average over the NSP is controlled at the nominal level.
Because the sample size is chosen to achieve a target average power while controlling the average type I error rate at the nominal level, this approach can be viewed as a hybrid frequentist--Bayesian design \citep{brutti-BayesianfrequentistSampleSize-2014}.
It provides optimized analysis prior parameters, such as the power parameter in the power prior and the weight parameter in the robust mixture prior, to achieve the target Bayesian power while controlling the Bayesian type I error rate at the nominal level.

Although the framework provides a practical approach to Bayesian sample size determination, some issues remain in the specification of the sampling and analysis priors.
In addition, the framework has not been extended to hybrid-controlled trials.

Regarding the sampling priors, because the NSP is constructed from external data, this relaxation may not actually be achieved, depending on the external data.
In the default specification of Psioda and Ibrahim, the NSP is the posterior distribution derived from external data restricted to the null hypothesis space.
When the external data exhibit a large treatment effect or sample size, the posterior distribution derived from the external data is concentrated within the alternative hypothesis space and the NSP consequently degenerates toward the boundary of the null hypothesis.
As discussed in Section \ref{sec3}, the Bayesian type I error rate in this case approaches the frequentist type I error rate at the boundary of the hypothesis space, so that borrowing external data may not reduce the required sample size compared with a frequentist design without borrowing.

Regarding the analysis prior, Psioda and Ibrahim used several options including a two-component robust mixture prior.
Although the second component of a robust mixture prior is generally specified as a weakly informative prior \citep{schmidli-RobustMetaanalyticpredictivePriors-2014}, Psioda and Ibrahim mainly presented simulation results using a second component centered at the boundary of the hypothesis space with the same amount of information as the external data.
This second component is equivalent to hypothetical data of the same size as the external data, assuming no treatment effect.
When the current trial data are inconsistent with the external data (i.e., prior-data conflict), this second component may bias the treatment effect estimate toward the null hypothesis space.
We therefore evaluate the impact of the choice of the mixture prior on the estimation of the treatment effect.
The details are discussed in Section \ref{sec3}.

Regarding practical applicability, the original method proposed by Psioda and Ibrahim was primarily designed for single-arm or two-arm trials in which external data are available for both arms.
Their framework is not directly applicable to scenarios in which only external control data are used for current two-arm trials (i.e., hybrid-controlled trials).
However, in clinical practice, borrowing external control data for a current two-arm trial is an attractive strategy because it remains applicable even when the experimental treatments in the external and current trials differ, provided that the inclusion criteria and patient backgrounds are consistent.
Additionally, borrowing external control data allows for an increased allocation ratio to the treatment group by minimizing the size of the concurrent control arm.
Therefore, extending the framework to accommodate these scenarios is important for clinical applications.

In this study, we address these issues as follows.
First, we propose an alternative specification of the NSP to achieve the relaxation of strict type I error control that Psioda and Ibrahim intended.
Second, we evaluate how the choice of mixture prior as the analysis prior affects the estimation of the treatment effect.
Finally, we extend the framework to accommodate hybrid-controlled trials.
We applied the proposed method to simulation studies and a case study to evaluate the required sample size and the estimation performance of the treatment effect.

The remainder of this paper is organized as follows.
In Section \ref{sec2}, we describe the Bayesian analysis method and sample size determination framework.
Section \ref{sec3} reviews the sampling and analysis priors established by Psioda and Ibrahim, discusses when the intended relaxation is not achieved, and presents the proposed specification.
In Section \ref{sec4}, we extend the framework to hybrid-controlled trial settings.
In Section \ref{sec5}, we conduct simulation studies to compare the required sample size and the estimation performance of the existing method by Psioda and Ibrahim and the proposed method.
In Section \ref{sec6}, the proposed methodology is applied to a clinical trial design for cutaneous lupus erythematosus to demonstrate its practical utility.
Finally, we discuss the findings of our numerical experiments and real-world applications in Section \ref{sec7}.

\section{Bayesian analysis and sample size determination for single-arm trials} \label{sec2}
In this section, we define the clinical trial setting and the notation used in this study.
We then describe the general framework for Bayesian analysis and simulation-based sample size determination.

\subsection{Specifications and notation}
We consider a single-arm clinical trial with a continuous outcome.
We assume that there is one available external dataset collected from a previous trial that evaluated the same treatment in a comparable population.
Let $\theta$ denote the treatment effect parameter of interest with parameter space $\Theta$.
We aim to test the following one-sided hypotheses.
The null hypothesis $\text{H}_0$ and the alternative hypothesis $\text{H}_1$ are defined as follows:
\begin{align*}
	\text{H}_0: \theta \in \Theta_0 = \{\theta : \theta \ge\lambda\} \quad \text{versus} \quad \text{H}_1: \theta\in \Theta_1 = \{\theta : \theta < \lambda\},
\end{align*}
where $\lambda$ is a clinically meaningful threshold.
Notably, while we formulate the problem as a superiority test (where lower values indicate better efficacy) to align with our motivating example, the framework can be generalized to other hypothesis settings.

Let $N_\text{E}$ and $N_\text{C}$ be the sample sizes of the external and current trial data, respectively.
We assume that outcomes of the external data $y_{\text{E}, i_\text{E}}$ and the current data $y_{\text{C}, i_\text{C}}$ follow normal distributions with a known common variance $\tau^2$:
\begin{align}
	y_{\text{E}, i_\text{E}} & \overset{\text{i.i.d.}}{\sim} N(\theta, \tau^2) \quad (i_\text{E} = 1,\ldots,N_\text{E}), \label{dist_ex_data}  \\
	y_{\text{C}, i_\text{C}} & \overset{\text{i.i.d.}}{\sim} N(\theta, \tau^2) \quad (i_\text{C} = 1,\ldots,N_\text{C}). \label{dist_new_data}
\end{align}
Let $\boldsymbol{y}_\text{E} = (y_{\text{E},1}, \dots, y_{\text{E},N_\text{E}})^\top$ and $\boldsymbol{y}_\text{C} = (y_{\text{C},1}, \dots, y_{\text{C},N_\text{C}})^\top$ denote the vectors of observed outcomes.

\subsection{Bayesian analysis}
In the Bayesian framework, inference about the treatment effect $\theta$ is based on the posterior distribution $\pi(\theta \mid \boldsymbol{y}_\text{C}, \boldsymbol{y}_\text{E})$, which integrates the prior distribution with the observed data $\boldsymbol{y}_\text{C}$.
According to Bayes' theorem, the posterior is proportional to the product of the likelihood of the current data and the prior distribution informed by the external data:
\begin{align*}
	\pi (\theta\mid \boldsymbol y_\text{C}, \boldsymbol y_\text{E}) \propto L(\theta \mid \boldsymbol y_\text{C})\times\pi^{(a)}(\theta\mid \boldsymbol y_\text{E}),
\end{align*}
where $L(\theta \mid \boldsymbol y_\text{C})$ is the likelihood function based on the current data, and $\pi^{(a)}(\theta\mid \boldsymbol y_\text{E})$ is the analysis prior derived from the external data.
Under the normality assumption in \eqref{dist_new_data}, the likelihood is given by
\begin{align*}
	L(\theta \mid \boldsymbol y_\text{C}) = \prod_{i_\text{C}=1}^{N_\text{C}} \frac{1}{\sqrt{2\pi\tau^2}}\exp\left\{-\frac{(y_{\text{C},i_\text{C}}-\theta)^2}{2\tau^2}\right\}.
\end{align*}
The choice of the analysis prior $\pi^{(a)}(\theta \mid \boldsymbol{y}_\text{E})$ is a key component of the design and is discussed in Section \ref{sec3}.
Based on the derived posterior distribution, we evaluate the treatment effect.
Specifically, we reject the null hypothesis $\text{H}_0$ if the posterior probability $T(\boldsymbol{y}_\text{C}) = \Pr(\theta \in \Theta_1\mid \boldsymbol{y}_\text{C}, \boldsymbol y_\text{E})$ exceeds the prespecified threshold $\phi \in (0,1)$ (e.g., $\phi=0.975$).

\subsection{Simulation-based sample size determination}
Following \citet{psioda-BayesianClinicalTrial-2019}, we employ a simulation-based approach to determine the required sample size $N_\text{C}$ for the current trial.
As current data are unobserved during the design stage of clinical trials, we introduce the sampling prior $\pi^{(s)}(\theta)$ to generate simulated data for the current trial $\boldsymbol{\tilde{y}}_\text{C}=(\tilde y_{\text{C},1},\dots, \tilde y_{\text{C},N_\text{C}})^\top$.
As with $\boldsymbol y_\text{C}$, we assume that the generated data $\tilde y_{\text{C}, i_\text{C}}$ follow the distribution in \eqref{dist_new_data}.
\begin{align*}
	\tilde y_{\text{C}, i_\text{C}} \stackrel{\text{i.i.d.}}{\sim} N(\theta, \tau^2)  \quad (i_\text{C} = 1,\ldots,N_\text{C}).
\end{align*}
The sampling prior $\pi^{(s)}(\theta)$ is defined under both the null and alternative hypotheses, referred to as the null sampling prior (NSP) $\pi^{(s)}_0(\theta)$ and the alternative sampling prior (ASP) $\pi^{(s)}_1(\theta)$, respectively.
If the hypothesis $\text{H}_\ell \, (\ell \in \{0,1\})$ is true, the marginal density function $f_\ell(\boldsymbol{\tilde{y}}_\text{C})$ of $\boldsymbol{\tilde{y}}_\text{C}$ under $\text{H}_\ell$ is given by
\begin{align}\label{generated new data}
	\boldsymbol{\tilde{y}}_\text{C} \sim f_\ell(\boldsymbol{\tilde{y}}_\text{C}) = \int f_N(\boldsymbol{\tilde{y}}_\text{C} \mid \theta, \tau^2) \pi^{(s)}_\ell(\theta) \, d\theta,
\end{align}
where $f_N(\boldsymbol{\tilde{y}}_\text{C} \mid \theta, \tau^2)$ is the joint normal density function of $\boldsymbol{\tilde{y}}_\text{C}$ with mean $\theta$ and variance $\tau^2$.

To determine the required sample size $N_\text{C}$ using the generated data $\boldsymbol{\tilde{y}}_\text{C}$ based on \eqref{generated new data}, we calculate the posterior distribution $\pi(\theta \mid \boldsymbol{\tilde{y}}_\text{C}, \boldsymbol{y}_\text{E})$ and evaluate the treatment effect.
Specifically, for the simulated data $\boldsymbol{\tilde{y}}_\text{C}$, we reject the null hypothesis $\text{H}_0$ if the posterior probability $T(\boldsymbol{\tilde {y}}_\text{C})$ exceeds a prespecified threshold $\phi$.

Given $\boldsymbol{\tilde{y}}_\text{C}$ in \eqref{generated new data}, the power function is defined as the expectation of the rejection indicator with respect to the marginal distribution of $\boldsymbol{\tilde{y}}_\text{C}$:
\begin{equation} \label{power_function}
	\begin{split}
		\beta_\ell & = \mathrm E_{\pi^{(s)}_\ell}[\mathds{1} \{T(\boldsymbol{\tilde {y}}_\text{C}) \ge \phi\}]                                                                                                                              \\
		           & = \int \mathds{1} \{T(\boldsymbol{\tilde {y}}_\text{C}) \ge \phi\} f_\ell(\boldsymbol{\tilde {y}}_\text{C}) \, d \boldsymbol{\tilde {y}}_\text{C}                                                                      \\
		           & = \int \mathds{1} \{T(\boldsymbol{\tilde {y}}_\text{C}) \ge \phi\} \left(\int f_N(\boldsymbol{\tilde{y}}_\text{C} \mid \theta, \tau^2) \pi^{(s)}_\ell(\theta) \, d\theta\right) \, d \boldsymbol{\tilde {y}}_\text{C},
	\end{split}
\end{equation}
where $\mathds{1}\{\cdot\}$ denotes the indicator function.
\begin{align*}
	\mathds{1}\{T(\boldsymbol{\tilde {y}}_\text{C})\ge\phi\} =
	\begin{cases}
		1 & \text{if } T(\boldsymbol{\tilde {y}}_\text{C})\ge\phi, \\
		0 & \text{if } T(\boldsymbol{\tilde {y}}_\text{C})<\phi.
	\end{cases}
\end{align*}
The Bayesian type I error rate ($\alpha$) and power ($1-\beta$) correspond to $\beta_0$ and $\beta_1$, respectively.

As the analytical derivation of the Bayesian type I error rate and power is generally impossible, these operating characteristics are calculated via Monte Carlo simulations for a fixed $N_\text{C}$ according to the following procedure:
\begin{enumerate}
	\item Sample $\theta$ from the sampling prior $\pi^{(s)}_\ell(\theta)$.
	\item Generate current data $\boldsymbol{\tilde{y}}_\text{C}$ based on the sampled $\theta$.
	\item Calculate $T(\boldsymbol{\tilde {y}}_\text{C}) = \Pr(\theta \in \Theta_1 \mid \boldsymbol{\tilde {y}}_\text{C}, \boldsymbol y_\text{E})$.
	\item Repeat steps 1--3 for a sufficiently large number of iterations ($N_{\text{sim}}$ times) and compute the average of $\mathds{1} \{T(\boldsymbol{\tilde {y}}_\text{C}) \ge \phi\}$ to estimate the Bayesian type I error rate or power.
\end{enumerate}
Through this procedure, $\alpha$ and $1-\beta$ for the specified sample size $N_\text{C}$ are approximately calculated.
The required sample size $N_\text{C}$ is then determined as the minimum value that satisfies the criteria for $\alpha$ and $1-\beta$.

\section{Existing and proposed specifications}\label{sec3}

\subsection{Sampling prior specifications}
\subsubsection{Existing specifications by Psioda and Ibrahim}
Psioda and Ibrahim constructed a sampling prior $\pi^{(s)}_\ell(\theta)$ ($\ell \in \{0,1\}$) based on the posterior distribution of the external data $\pi(\theta \mid \boldsymbol{y}_\text{E})$.
$\pi(\theta \mid \boldsymbol y_\text{E})$ is derived using Bayes' theorem, as follows:
\begin{align*}
	\pi(\theta \mid \boldsymbol y_\text{E}) \propto L(\theta \mid \boldsymbol y_\text{E})\times \pi_0(\theta),
\end{align*}
where $L(\theta \mid \boldsymbol y_\text{E})$ is the likelihood function based on external data, and $\pi_0(\theta)$ is a prior of $\pi(\theta \mid \boldsymbol y_\text{E})$.
Under the assumption that the external data follow the distribution in \eqref{dist_ex_data}, and using a non-informative prior, the posterior distribution $\pi(\theta \mid \boldsymbol y_\text{E})$ is given by
\begin{align}\label{sp}
	\pi(\theta \mid \boldsymbol y_\text{E}) = f_N\left(\theta \relmiddle| \bar y_\text{E}, \frac{\tau^2}{N_\text{E}}\right),
\end{align}
where $\bar y_\text{E} = \sum_{i_\text{E}=1}^{N_\text{E}} y_{\text{E},i_\text{E}}/N_\text{E}$.
In the default specification, the NSP and ASP are the posterior distributions $\pi(\theta \mid \boldsymbol y_\text{E})$ restricted to $\Theta_0$ and $\Theta_1$, respectively.
As an option, they also proposed truncating the tails of these distributions when stakeholders consider some parameter values to be implausible.
The truncated NSP and ASP are defined as follows:
\begin{align*}
	\pi^{(s)}_0(\theta) & = \frac{\pi(\theta \mid \boldsymbol y_\text{E})}{\int_{\tilde{\Theta}_0} \pi(\theta \mid \boldsymbol y_\text{E})\,d\theta} \mathds{1}\{\theta \in \tilde{\Theta}_0\}, \\
	\pi^{(s)}_1(\theta) & = \frac{\pi(\theta \mid \boldsymbol y_\text{E})}{\int_{\tilde{\Theta}_1} \pi(\theta \mid \boldsymbol y_\text{E})\,d\theta} \mathds{1}\{\theta \in \tilde{\Theta}_1\}.
\end{align*}
Here, $\tilde{\Theta}_\ell$ is the truncated parameter space, defined as
\begin{align*}
	\tilde{\Theta}_\ell = \left\{ \theta \in \Theta_\ell : \pi(\theta \mid \boldsymbol y_\text{E}, \text{H}_\ell) \ge \frac{1}{K} \max_{\theta \in \Theta_\ell} \pi(\theta \mid \boldsymbol y_\text{E},
	\text{H}_\ell) \right\},
\end{align*}
where $K$ is a prespecified constant.
In their simulation study, they used $K=2$, and we use this specification as the existing method in Section \ref{sec5}.
Figure \ref{fig: sp}(a) illustrates the sampling prior proposed by Psioda and Ibrahim with $K=2$.

\subsubsection{Proposed specifications}
Their sampling prior aims to relax the strict control of the type I error rate.
Such relaxation is necessary for borrowing external data to be beneficial because controlling the Bayesian type I error rate at the nominal level is difficult when external data exhibiting a treatment effect are incorporated.
As also noted by \citet{kopp-schneider-PowerGainsUsing-2020}, the type I error rate can inflate as external data are borrowed, which may lead to abandoning borrowed information if we attempt to control the Bayesian type I error rate at the same strict level as in frequentist analysis.
In addition, \citet{kopp-schneider-PowerGainsUsing-2020} argued that specifying the sampling prior as a point mass is a pessimistic and arbitrary choice that fails to account for the uncertainty of the parameter.

Under the specification of Psioda and Ibrahim, the NSP is constructed by restricting the posterior distribution $\pi(\theta \mid \boldsymbol y_\text{E})$ derived from the external data to $\Theta_0$.
When the external data exhibit a large treatment effect or a large sample size, the NSP tends to degenerate toward the boundary $\lambda$.
Although evaluating the Bayesian type I error rate using a sampling prior was originally intended to relax strict frequentist type I error control, this degeneration essentially reverts the evaluation back to a strict point-mass scenario, failing to achieve the intended relaxation.
When the NSP is specified as a point mass at $\lambda$, and no information is borrowed from the external data, the Bayesian type I error rate theoretically coincides with the type I error rate in frequentist hypothesis testing for $\theta = \lambda$ (see Appendix \ref{app: bayesian_freq_equivalence} for details).
Consequently, it becomes difficult to control the Bayesian type I error rate at the nominal level when external data exhibiting a treatment effect are incorporated.

Therefore, we propose an alternative way to elicit the NSP.
We specify the NSP as a truncated normal distribution centered at $\lambda$ to prevent incorporation of external data that are excessively consistent with the alternative hypothesis when evaluating the design under the null scenario, and to control the Bayesian type I error rate under more relaxed and practically feasible conditions.
Our proposed sampling priors are defined as follows:
\begin{align*}
	\pi^{(s)}_0(\theta) & = \frac{f_N\left(\theta\relmiddle| \lambda, \frac{\tau^2}{N_\text{E}}\right)}{\int_{\Theta_0} f_N\left(\theta\relmiddle| \lambda, \frac{\tau^2}{N_\text{E}}\right) d\theta} \mathds{1}\{\theta \in \Theta_0\},               \\
	\pi^{(s)}_1(\theta) & = \frac{f_N\left(\theta\relmiddle|\bar y_\text{E}, \frac{\tau^2}{N_\text{E}}\right)}{\int_{\Theta_1} f_N\left(\theta\relmiddle|\bar y_\text{E}, \frac{\tau^2}{N_\text{E}}\right) d\theta} \mathds{1}\{\theta \in \Theta_1\}.
\end{align*}
Here, the NSP is specified as a normal distribution centered at the threshold $\lambda$ of the null hypothesis.
This proposed NSP is the posterior distribution that would be constructed from hypothetical external data with the sample mean equal to $\lambda$ and with the same sample size $N_\text{E}$ as the actual external data.
As in the default specification of Psioda and Ibrahim, no additional tail truncation (i.e., $K=\infty$) is applied to our sampling priors.
This is because, although the parameter space of each hypothesis is restricted, the true treatment effect in the current trial may lie outside the truncated parameter space $\tilde \Theta_{\ell}$.
In this case, the sampling prior does not cover the true treatment effect, which may result in lower power than the target level.

\begin{figure}[H]
	\centering
	\includegraphics[width = 0.9\linewidth]{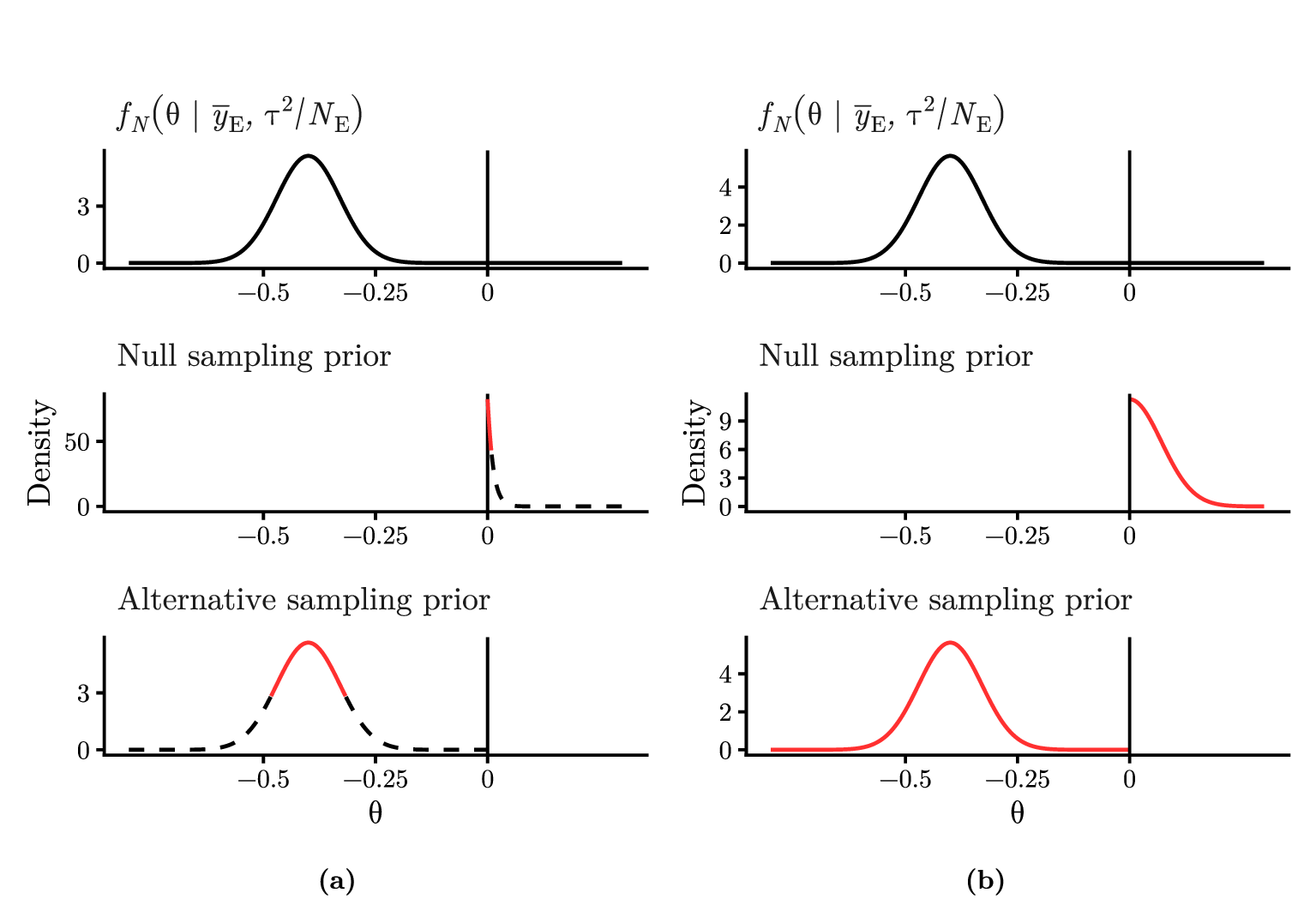}
	\caption{Comparison of sampling priors with $\lambda=0$, $\bar y_\text{E}=-0.4$ and $\tau^2/N_\text{E}=0.02$. The red solid lines represent the region of the sampling priors. (a) Existing sampling prior by Psioda and Ibrahim ($K=2$). (b) Proposed sampling prior.}
	\label{fig: sp}
\end{figure}

The sampling priors of the proposed specification are shown in Figure \ref{fig: sp}(b).

\subsection{Analysis prior specifications}
\subsubsection{Existing specifications by Psioda and Ibrahim}
Psioda and Ibrahim employed a robust mixture prior with two components as an option for the analysis prior.
A robust mixture prior with two components is expressed as follows:
\begin{align}\label{analytic_prior_ge}
	\pi^{(a)}(\theta\mid \boldsymbol y_\text{E}) = w \times f_1(\theta \mid \boldsymbol y_\text{E}) + (1-w) \times f_2(\theta),
\end{align}
where $w \in [0,1]$ is a weight parameter.
The first component $f_1(\theta \mid \boldsymbol y_\text{E})$ is set as the posterior distribution of the external data $\pi(\theta \mid \boldsymbol y_\text{E})$ in \eqref{sp}, which fully incorporates information from the external data.
The parameter $w$ controls the degree of borrowing from the external data.
The weight $w$ is updated depending on the heterogeneity between the external and current trial data.

The posterior distribution $\pi(\theta \mid \boldsymbol y_\text{C}, \boldsymbol y_\text{E})$ is then derived using Bayes' theorem as follows:
\begin{align*}
	\pi(\theta \mid \boldsymbol y_\text{C}, \boldsymbol y_\text{E}) & \propto L(\theta \mid \boldsymbol y_\text{C})\times\pi^{(a)}(\theta\mid \boldsymbol y_\text{E})                                      \\
	                                                                & = w' \times f_1(\theta \mid \boldsymbol y_\text{C}, \boldsymbol y_\text{E}) + (1-w') \times f_2(\theta \mid \boldsymbol y_\text{C}),
\end{align*}
where $w'$ is the updated (posterior) weight parameter.
\begin{align}\label{posterior w calc}
	w' = \frac{w m_1(\boldsymbol y_\text{C})}{w m_1(\boldsymbol y_\text{C})+(1-w) m_2(\boldsymbol y_\text{C})},
\end{align}
where $m_1(\boldsymbol y_\text{C})$ and $m_2(\boldsymbol y_\text{C})$ are marginal likelihoods with respect to $f_1(\theta \mid \boldsymbol y_\text{E}), f_2(\theta)$, respectively, and are defined as follows:
\begin{align*}
	m_1(\boldsymbol y_\text{C}) & = \int L(\theta \mid \boldsymbol y_\text{C}) f_1(\theta \mid \boldsymbol y_\text{E}) \, d\theta, \\
	m_2(\boldsymbol y_\text{C}) & = \int L(\theta \mid \boldsymbol y_\text{C}) f_2(\theta) \, d\theta.
\end{align*}
A detailed derivation of the posterior distribution and updated weight parameters is provided in Appendix \ref{app: calc_posterior_ap_and_w}; see also \citet{ratta-InterplayPriorWeight-2026}.

The analysis priors specified by Psioda and Ibrahim are as follows:
\begin{align}\label{ex_analytic_prior}
	\pi^{(a)}(\theta\mid \boldsymbol y_\text{E}) = w \times f_N \left(\theta \relmiddle| \bar y_\text{E}, \frac{\tau^2}{N_\text{E}} \right) + (1-w) \times f_N \left(\theta \relmiddle| \lambda, \frac{\tau^2}{N_\text{E}} \times k \right),
\end{align}
where $k \in \{0.5,1.0,\dots,10.0\}$ is a constant.
The first component $f_N \left(\theta \mid \bar y_\text{E}, \tau^2/N_\text{E} \right)$ fully incorporates information from the external data, which is the same as in \eqref{sp}.

\subsubsection{Proposed specifications}
The second component in \eqref{ex_analytic_prior} is set as a highly informative prior centered at the threshold $\lambda$, particularly when $k$ is small.
The second component, $f_2(\theta)$, in \eqref{analytic_prior_ge} is generally set as a weakly informative prior to mitigate the potential impact of prior-data conflict \citep{schmidli-RobustMetaanalyticpredictivePriors-2014, Weru05052026}.
This is because when there is heterogeneity between the external and current trial data, $w$ is updated to a small value in \eqref{posterior w calc}, which leads to less borrowing from the external data if the second component is a weakly informative prior.
However, for example, when $k=1$ in \eqref{ex_analytic_prior}, which is the primary setting used in their simulation study, the second component contains information equivalent to a hypothetical dataset of size $N_\text{E}$ with mean $\lambda$.
When this analysis prior is used, the posterior distribution becomes a mixture of posterior distributions based on the external and hypothetical data with mean $\lambda$, which may lead to an underestimation of the treatment effect and complicate the interpretation of the posterior distribution.
Therefore, we set the second component as a weakly informative prior with a large variance, and apply the sample size determination method to the simulation studies to evaluate the performance of the following analysis prior:
\begin{align}\label{proposed_analytic_prior}
	\pi^{(a)}(\theta\mid \boldsymbol y_\text{E}) = w \times f_N \left(\theta \relmiddle| \bar y_\text{E}, \frac{\tau^2}{N_\text{E}} \right) + (1-w) \times f_N \left(\theta \mid \lambda, \tau^2 \right),
\end{align}
where $f_N \left(\theta \mid \lambda, \tau^2 \right)$ is a unit-information prior.
The second component is specified as a weakly informative prior to mitigate the potential impact of prior-data conflict when the external and current trial data are heterogeneous.
A completely non-informative prior (e.g., with variance approaching infinity) is generally inappropriate in a robust mixture prior because it may lead to posterior weights $w'$ degenerating to 1 unless the prior weight $w$ is set to be extremely small, thereby effectively eliminating the influence of the second component (see Appendix \ref{app: no_info} for details).
Therefore, we specify the second component as a unit-information prior, which is a commonly used weakly informative prior in the context of robust mixture priors \citep{schmidli-RobustMetaanalyticpredictivePriors-2014, Weru05052026}.
A recent study has highlighted that the posterior distribution of a robust mixture prior depends not only on $w$ but also on the specification of the variance of the second component \citep{ratta-InterplayPriorWeight-2026}.
In this study, the variance of the second component is fixed a priori, and $w$ is calibrated to achieve the desired operating characteristics for the sample size determination.

\section{Sample size determination for hybrid-controlled trials}\label{sec4}
In this section, we extend this sample size determination framework to hybrid-controlled trials.
We assume that external control data are available, and the objective is to evaluate the treatment effect by comparing the current treatment arm with the control arm using both the external and current control data.
Let $\theta_\text{C}$ and $\theta_\text{T}$ denote the treatment effect parameters for the control and treatment arms, respectively.
The hypotheses of interest are as follows:
\begin{align*}
	\text{H}_0: \delta \ge \lambda \quad \text{versus} \quad \text{H}_1: \delta < \lambda,
\end{align*}
where $\delta = \theta_\text{T} - \theta_\text{C}$, and $\lambda$ is a clinically meaningful threshold for the treatment effect difference.
The sampling and analysis prior specifications can be similarly constructed to those in the single-arm setting with appropriate modifications to account for both the control and treatment arms, where the parameter of interest is the difference $\delta$.
We denote the outcomes for the external control arm as $\boldsymbol{y}_\text{EC}$, and the observed outcomes for the control and treatment arms in the current trial as $\boldsymbol{y}_\text{CC}$ and $\boldsymbol{y}_\text{CT}$, respectively.
Each observed outcome follows a normal distribution with a known variance as follows:
\begin{align}
	y_{\text{EC}, i_\text{EC}} & \overset{\text{i.i.d.}}{\sim} N(\theta_\text{C}, \tau^2) \quad (i_\text{EC} = 1,\ldots,N_\text{EC}), \label{dist_ec_data} \\
	y_{\text{CC}, i_\text{CC}} & \overset{\text{i.i.d.}}{\sim} N(\theta_\text{C}, \tau^2) \quad (i_\text{CC} = 1,\ldots,N_\text{CC}),\notag                \\
	y_{\text{CT}, i_\text{CT}} & \overset{\text{i.i.d.}}{\sim} N(\theta_\text{T}, \tau^2) \quad (i_\text{CT} = 1,\ldots,N_\text{CT}), \notag
\end{align}
where $\tau^2$ is the known variance, and $N_\text{EC}$, $N_\text{CC}$, and $N_\text{CT}$ are the sample sizes for the external control data, the control arm in the current trial, and the treatment arm in the current trial, respectively.

Then, we define the sampling priors for $\theta_\text{C}$ and $\theta_\text{T}$ under the null and alternative hypotheses based on the external control and current trial data, with constraints reflecting the hypotheses.
As in the single-arm setting, when the external control data follow the distribution in \eqref{dist_ec_data}, the sampling prior for $\theta_\text{C}$ is specified as a normal distribution with mean $\bar{y}_\text{EC}$ and variance $\tau^2/N_\text{EC}$.

Under the null hypothesis $\text{H}_0$, we specify the sampling prior for $\theta_\text{T}$ as a normal distribution with mean $\bar{y}_\text{EC}$ and variance $\tau^2/N_\text{EC}$.
To reflect the assumption that the treatment is ineffective or inferior under $\text{H}_0$, i.e., $\delta = \theta_\text{T} - \theta_\text{C} \ge \lambda$, we constrain the parameter space:
\begin{align*}
	\pi_0^{(s)}(\theta_\text{C}) & = f_N\left(\theta_\text{C} \relmiddle| \bar y_\text{EC}, \frac{\tau^2}{N_\text{EC}}\right),                                                                                \\
	\pi_0^{(s)}(\theta_\text{T}) & = f_N\left(\theta_\text{T} \relmiddle| \bar y_\text{EC}, \frac{\tau^2}{N_\text{EC}}\right) \quad \text{s.t.} \quad \delta = \theta_\text{T} - \theta_\text{C} \ge \lambda.
\end{align*}

Under the alternative hypothesis $\text{H}_1$, the treatment effect for the treatment arm is smaller than that of the control arm, i.e., the treatment group shows greater efficacy.
Thus, we specify the sampling prior for $\theta_\text{T}$ as a normal distribution with mean $\bar{y}_\text{EC} + \delta_0$ and variance $\tau^2/N_\text{EC}$, where $\delta_0$ represents the target treatment effect assumed for the treatment in the current trial.
To reflect the assumption that the treatment is superior under $\text{H}_1$, i.e., $\delta = \theta_\text{T} - \theta_\text{C} < \lambda$, we constrain the parameter space:
\begin{align*}
	\pi_1^{(s)}(\theta_\text{C}) & = f_N\left(\theta_\text{C} \relmiddle| \bar y_\text{EC}, \frac{\tau^2}{N_\text{EC}}\right),                                                                                         \\
	\pi_1^{(s)}(\theta_\text{T}) & = f_N\left(\theta_\text{T} \relmiddle| \bar y_\text{EC} + \delta_0, \frac{\tau^2}{N_\text{EC}}\right) \quad \text{s.t.} \quad \delta = \theta_\text{T} - \theta_\text{C} < \lambda.
\end{align*}

The analysis priors for $\theta_\text{C}$ and $\theta_\text{T}$ are defined as follows:
\begin{align*}
	\pi^{(a)}(\theta_\text{C}\mid \boldsymbol y_\text{EC}) & = w \times f_N \left(\theta_\text{C} \relmiddle| \bar y_\text{EC}, \frac{\tau^2}{N_\text{EC}} \right) + (1-w) \times f_N \left(\theta_\text{C} \mid \bar y_\text{EC}, \tau^2 \right), \\
	\pi^{(a)}(\theta_\text{T})                             & = f_N \left(\theta_\text{T} \mid 0, 100^2 \right).
\end{align*}
The analysis prior for $\theta_\text{C}$ is a robust mixture prior incorporating information from the external control data.
As discussed in Section \ref{sec3}, the second component of the analysis prior is set as a unit-information prior to mitigate prior-data conflict.
The analysis prior for $\theta_\text{T}$ is non-informative, reflecting the lack of prior information about the treatment effect in the treatment arm.

In the hybrid-controlled trial setting, we assume that the external and current control data do not conflict, and the heterogeneity between the external and current control data is handled through the analysis prior.
Accordingly, the sampling prior for $\theta_\text{C}$ is specified with mean $\bar{y}_\text{EC}$ and variance $\tau^2/N_\text{EC}$ under both the null and alternative hypotheses.
This implies that the heterogeneity between the external and current control data is not considered during the design stage of clinical trials.
Therefore, when $w$ is optimized by grid search as in the single-arm setting, the optimal $w$ tends to be close to 1.
As a result, the required sample size may become unrealistically small because the posterior variance is almost the same in full borrowing, satisfying the nominal criteria for the Bayesian type I error and power.
To address this issue, we specify $w$ so that the total information in the external and current control arms becomes comparable to that in the current treatment arm.
The prior information in the control arm is quantified as the effective sample size (ESS) of the analysis prior $\pi^{(a)}(\theta_\text{C}\mid \boldsymbol y_\text{EC})$.
The procedure for determining $w$ is as follows:
\begin{enumerate}
	\item Calculate the $\mathrm{ESS}(w)$ of the analysis prior $\pi^{(a)}(\theta_\text{C}\mid \boldsymbol y_\text{EC})$ for each $w \in \{0, 0.001, \dots, 1\}$.
	\item Identify the value of $w$ as follows:
	      \begin{align*}
		      w = \arg\max_{w} \left\{ \text{ESS}(w) : N_\text{D} \le \text{ESS}(w) < N_\text{D}+1 \right\},
	      \end{align*}
	      where $N_\text{D} = N_\text{CT} - N_\text{CC}$.
\end{enumerate}
If no $w$ satisfies this condition, we set $w=1$ when $\text{ESS}(w)$ does not reach $N_\text{D}$ even at $w=1$ (e.g., owing to the limited sample size of the external control data), and $w=0$ when $\text{ESS}(0) \ge N_\text{D} + 1$, that is, when the target information is already attained without borrowing.

Using the specified sampling and analysis priors, we can derive the posterior distribution of $\theta_\text{C}$ and $\theta_\text{T}$ and then evaluate the treatment effect difference $\delta$.
The Bayesian type I error rate and power can be calculated using Monte Carlo simulations as described in the single-arm setting in Section \ref{sec2}.

This procedure assumes that the current control arm is smaller than the treatment arm, and aims to borrow external control data to make the total control information comparable to the treatment information.
This procedure can also be applied when a larger amount of control information is desired, for example, when the current trial is balanced or when researchers wish to use more of the available external control data, by replacing $N_\text{D}$ with $\rho N_\text{CT} - N_\text{CC}$, where $\rho\, (>0)$ is a prespecified ratio of the total control information, $\mathrm{ESS}(w) + N_\text{CC}$, to the treatment information, $N_\text{CT}$.
A larger $\rho$ increases the gain from borrowing external data, but also increases the potential bias when the external data and current control data are heterogeneous.
Therefore, $\rho$ should be chosen in the context of each trial.

\section{Simulation studies}\label{sec5}
We conducted simulation studies to determine the required sample sizes for both single-arm and hybrid-controlled trials.
Furthermore, for single-arm trials, we compared the performance of treatment effect estimation between the existing method by Psioda and Ibrahim and our proposed method.
Note that the existing and proposed methods define the Bayesian type I error rate under different null sampling priors.
Therefore, the comparison of the required sample sizes is not a comparison under a common criterion, and should be interpreted as illustrating whether the relaxation is achieved.

\subsection{Simulation study for single-arm trials}
\subsubsection{Design}
We assumed a single-arm clinical trial with a continuous outcome, as described in Section \ref{sec2}.
We defined the null and alternative hypotheses as $\text{H}_0: \theta \ge 0$ and $\text{H}_1: \theta < 0$, respectively.
The known variance was set to $\tau^2 = 1$.
We assumed that the sample size of the external data was $N_\text{E}\in \{50, 200\}$, and the mean outcome of the external data was set to $\bar y_\text{E} \in \{-0.4, -0.6, -0.8, -1.0\}$.

\subsubsection{Evaluation metrics}
The primary metric was the required sample size, defined as the minimum sample size $N_\text{C}$ that simultaneously satisfied the Bayesian type I error rate ($\alpha$) and power ($1-\beta$) constraints under each specification.
For every fixed $N_\text{C}$, we performed a grid search for the weight parameter $w$ over the set $\{0, 0.01, \dots, 1\}$. We calculated the Bayesian type I error rate and power by using Monte Carlo simulations with $1,000,000$ iterations ($N_{\text{sim}}=1,000,000$).
We identified the combination $(N_\text{C}, w)$ that satisfied $\alpha \le 0.025$ and $1-\beta \ge 0.80$.
If multiple values of $w$ satisfied the conditions for a given $N_\text{C}$, we selected the value of $w$ that maximized statistical power.
The rejection criterion for the null hypothesis was a posterior probability of the alternative hypothesis satisfying $T(\boldsymbol{\tilde{y}}_\text{C}) \ge 0.975$.
As benchmarks, we calculated the sample size using the existing method by Psioda and Ibrahim and the standard frequentist method, which does not incorporate external data.

Using the sample size $N_\text{C}$ and weight $w$ determined by each method, we evaluated the operating characteristics of the treatment effect estimation.
We assumed that the true treatment effect of the current trial, $\theta$, ranged from $-1.0$ to $0.5$ in increments of $0.1$ (covering both the efficacy and null scenarios).
For each true $\theta$, we generated hypothetical current trial data $\boldsymbol{y}_\text{C}$ and derived the posterior distribution $\pi(\theta \mid \boldsymbol{y}_\text{C}, \boldsymbol{y}_\text{E})$ using the same analysis prior and weight $w$ determined in the design stage.

The null hypothesis rejection rate, posterior mean, estimation bias, and 95\% credible interval (CI) were calculated.
The posterior mean estimate $\hat{\theta}(\boldsymbol{y}_\text{C})$ is given by the expectation of the posterior distribution:
\begin{align*}
	\hat{\theta}(\boldsymbol y_\text{C}) & = \mathrm{E}[\theta \mid \boldsymbol y_\text{C}, \boldsymbol y_\text{E}]                     \\
	                                     & = \int \theta \, \pi(\theta \mid \boldsymbol y_\text{C}, \boldsymbol y_\text{E}) \, d\theta.
\end{align*}
The bias of the treatment effect estimation was defined as the difference between the expected posterior mean and the true parameter value:
\begin{align*}
	\text{Bias} & = \mathrm{E}_{\boldsymbol y_\text{C} \mid \theta}\left[\hat{\theta}(\boldsymbol y_\text{C})\right] - \theta                         \\
	            & = \int \hat{\theta}(\boldsymbol y_\text{C}) \, f_N(\boldsymbol y_\text{C} \mid \theta, \tau^2) \, d\boldsymbol y_\text{C} - \theta,
\end{align*}
where the expectation is taken over the distribution of the current data $\boldsymbol{y}_\text{C}$.
The 95\% credible interval $C$ was calculated as the Highest Posterior Density (HPD) interval.
The lower and upper bounds of the HPD interval, denoted by $C_\text{lower}$ and $C_\text{upper}$, are defined as
\begin{align*}
	C_\text{lower} & = \min \{\theta : \pi(\theta \mid \boldsymbol{y}_\text{C}, \boldsymbol{y}_\text{E}) \ge h\}, \\
	C_\text{upper} & = \max \{\theta : \pi(\theta \mid \boldsymbol{y}_\text{C}, \boldsymbol{y}_\text{E}) \ge h\},
\end{align*}
where $h$ is the threshold, such that $\int_C \pi(\theta \mid \boldsymbol{y}_\text{C}, \boldsymbol{y}_\text{E}) \, d\theta = 0.95$.

\subsubsection{Results}
Table \ref{tab: res_sim} summarizes the sample sizes required by each method.
The required sample sizes under the specifications of Psioda and Ibrahim were comparable to or slightly larger than those of the standard frequentist method, even though their specification intended to relax type I error control.
On the other hand, compared with both the existing method by Psioda and Ibrahim and the frequentist design, the proposed method reduced the required sample size, particularly when the external data suggested a weaker treatment effect (i.e., smaller absolute values of $\bar{y}_\text{E}$).
Comparing the cases in which the sample sizes of the external data were $N_\text{E}=50$ and $N_\text{E}=200$, the proposed method demonstrated more pronounced sample size reductions for $N_\text{E}=50$.
Regarding the optimal $w$, the proposed method tended to select smaller values than those of the existing method.

\begin{table}[H]
	\captionsetup{font=normalsize}
	\subcaptionsetup{font=normalsize}
	\centering
	\caption{Required sample sizes $N_\text{C}$, weights $w$, and frequentist required sample sizes $N_{\text{freq}}$}\label{tab: res_sim}
	\begin{subtable}{\textwidth}
		\caption{Required sample sizes and weights for $N_\text{E} = 50$}
		\centering
		\begin{tabular}{cccccc}
			\hline
			\multirow{2}{*}{$\bar y_\text{E}$} & \multirow{2}{*}{$N_{\text{freq}}$} & \multicolumn{2}{c}{Existing Method} & \multicolumn{2}{c}{Proposed Method}                       \\ \cline{3-6}
			                                   &                                    & $N_\text{C}$                        & $w$                                 & $N_\text{C}$ & $w$  \\ \hline
			$-$0.4                             & 50                                 & 52                                  & 0.59                                & 42           & 0.38 \\
			$-$0.6                             & 22                                 & 23                                  & 0.64                                & 19           & 0.35 \\
			$-$0.8                             & 13                                 & 13                                  & 0.68                                & 11           & 0.35 \\
			$-$1.0                             & 8                                  & 8                                   & 0.69                                & 7            & 0.36 \\
			\hline
		\end{tabular}
		\label{tab: NE50}
	\end{subtable}

	\vspace*{1.0cm}

	\begin{subtable}{\textwidth}

		\caption{Required sample sizes and weights for $N_\text{E} = 200$}
		\centering
		\begin{tabular}{cccccc}
			\hline
			\multirow{2}{*}{$\bar y_\text{E}$} & \multirow{2}{*}{$N_{\text{freq}}$} & \multicolumn{2}{c}{Existing Method} & \multicolumn{2}{c}{Proposed Method}                       \\ \cline{3-6}
			                                   &                                    & $N_\text{C}$                        & $w$                                 & $N_\text{C}$ & $w$  \\ \hline
			$-$0.4                             & 50                                 & 50                                  & 0.67                                & 42           & 0.19 \\
			$-$0.6                             & 22                                 & 23                                  & 0.71                                & 20           & 0.20 \\
			$-$0.8                             & 13                                 & 13                                  & 0.73                                & 12           & 0.22 \\
			$-$1.0                             & 8                                  & 8                                   & 0.74                                & 8            & 0.26 \\
			\hline
		\end{tabular}
		\label{tab: NE200}
	\end{subtable}
	\vspace{0.7cm}
\end{table}

\begin{figure}[H]
	\centering
	\includegraphics[width = 0.9\linewidth]{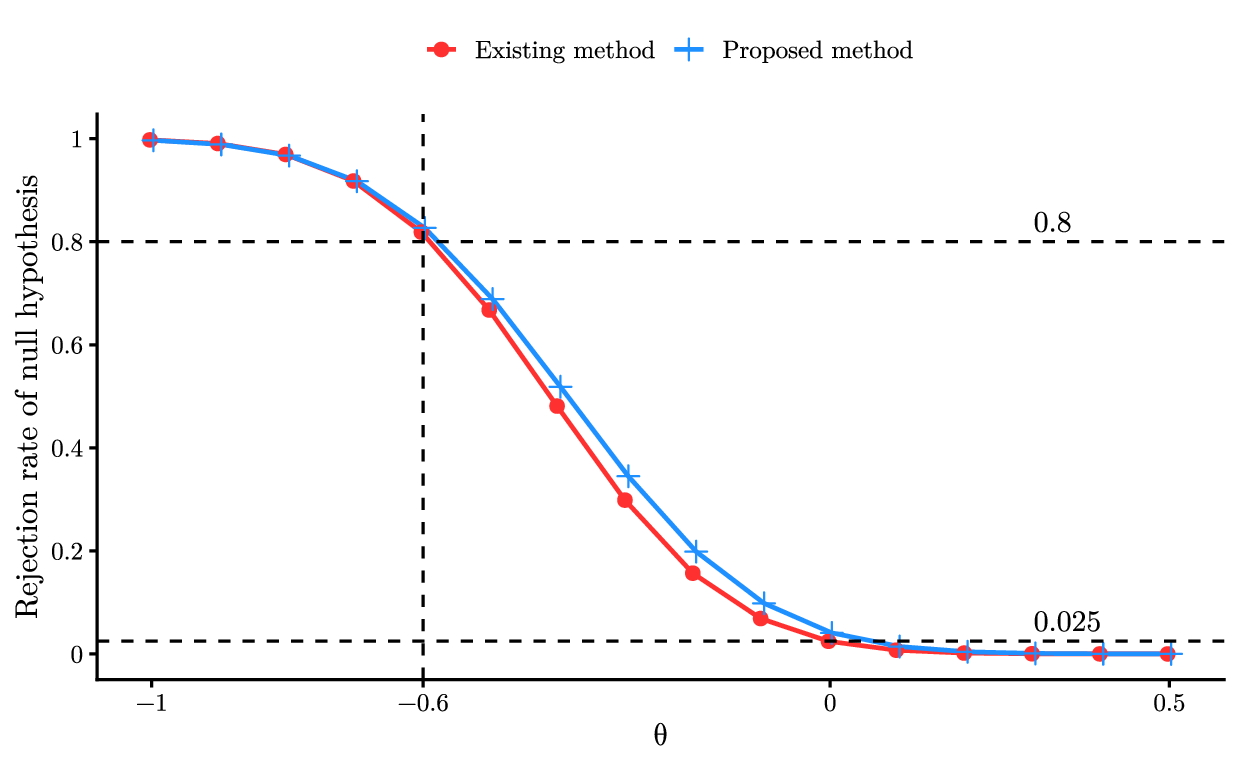}
	\caption{Rejection rates of the null hypothesis for $\bar y_\text{E} = -0.6$ and $N_\text{E} = 200$. The sample sizes $N_\text{C}$ for the existing and proposed methods were 23 and 20, respectively.}
	\label{figs: sim_power}
\end{figure}

\begin{figure}[H]
	\centering
	\includegraphics[width = 0.9\linewidth]{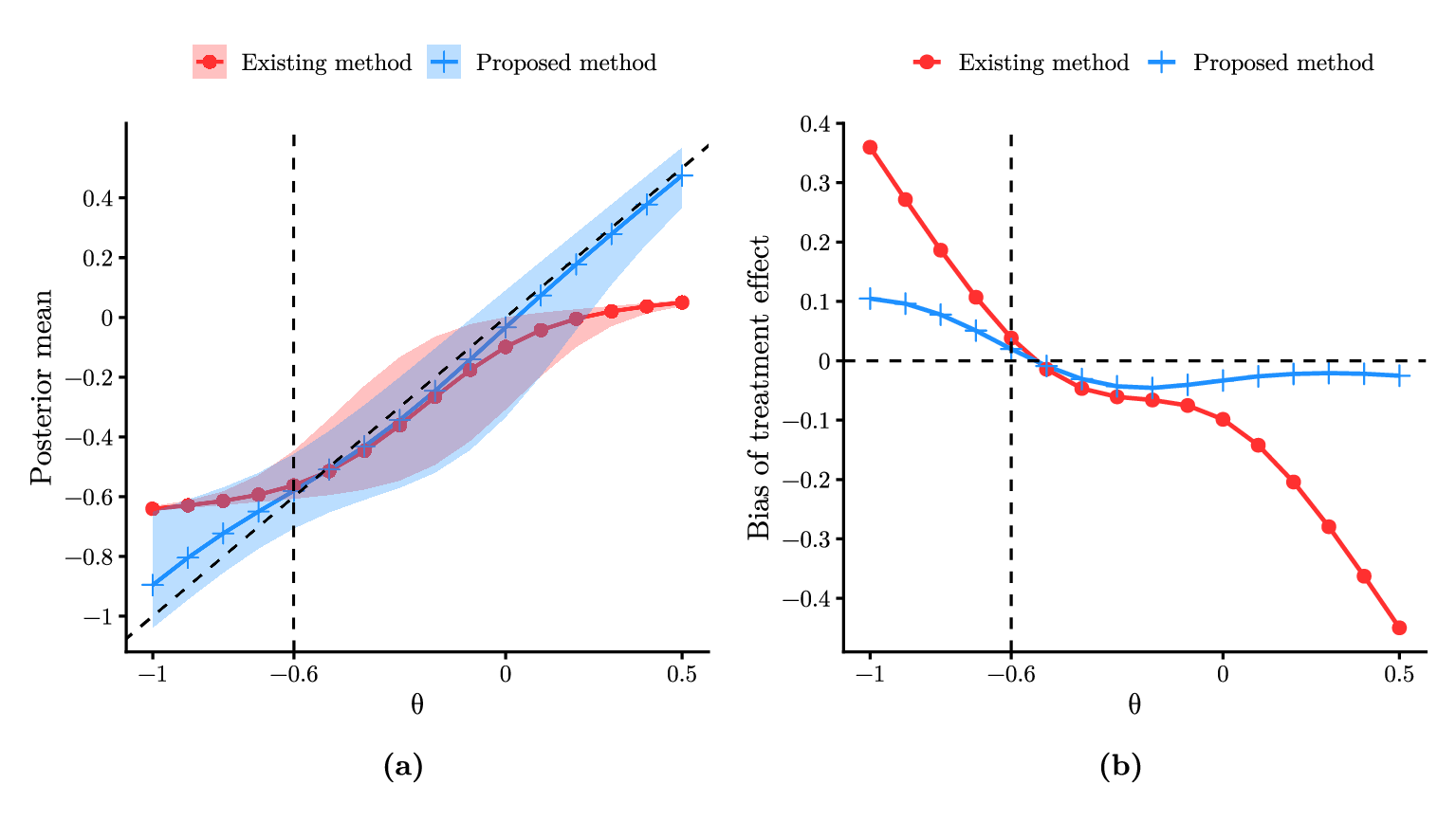}
	\caption{Posterior means, 95\% credible intervals (95\% CIs) and biases of the treatment effect estimates for $\bar y_\text{E} = -0.6$ and $N_\text{E} = 200$. The sample sizes $N_\text{C}$ for the existing and proposed methods were 23 and 20, respectively. (a) Posterior means and 95\% CIs. The solid lines represent the posterior means, and the bands represent the 95\% credible intervals. (b) Biases of treatment effect estimates.}
	\label{figs: sim_estimate_bias}
\end{figure}

Next, we evaluated the operating characteristics (rejection rates of the null hypothesis) and estimation performance (posterior means and biases) of the designs shown in Table \ref{tab: res_sim} assuming the true treatment effect $\theta \in \{-1.0,-0.9,\dots,0.5\}$.
Figures \ref{figs: sim_power} and \ref{figs: sim_estimate_bias} show the rejection rates of the null hypothesis, posterior means with 95\% credible intervals (CIs), and estimation biases for scenarios with $N_\text{E} = 200$ and $\bar{y}_\text{E} = -0.6$.
The results for the other scenarios showed similar trends and are provided in the Supplementary Materials. 
Based on Table \ref{tab: NE200}, the determined design parameters $(N_\text{C}, w)$ were $(23, 0.71)$ for the existing method and $(20, 0.20)$ for the proposed method.
The proposed method provided greater statistical power despite requiring a smaller sample size than the existing method (Figure \ref{figs: sim_power}).
Specifically, when the effect size of the current trial matched the external data ($\theta=-0.6$), the power was $0.819$ for the existing method and $0.827$ for the proposed method.
However, in the region where the treatment effect was small or worse (i.e., $\theta$ was greater than 0), the rejection rate of the proposed method was higher than that of the existing method.
At $\theta=0$ (the null hypothesis), the rejection rates (i.e., the frequentist type I error rates) for the existing and proposed methods were $0.025$ and $0.041$, respectively.
This exceedance of the nominal level at the boundary is a consequence of controlling the Bayesian type I error rate, which is discussed in Section \ref{sec7}.
Regarding the estimation performance, Figures \ref{figs: sim_estimate_bias}(a) and \ref{figs: sim_estimate_bias}(b) demonstrate that the proposed method produced smaller biases and avoided excessively narrow credible intervals compared to the existing method.
Notably, in the regions where $\theta \le -0.6$ or $\theta \ge 0$, the existing method produced very narrow 95\% CIs, and the posterior means shrank strongly toward $-0.6$ or $0$.
By contrast, the estimates from the proposed method closely followed the true treatment effect of the current data, indicating a reduced undesirable influence from both the external data and the second component of the analysis prior.

\subsection{Simulation study for hybrid-controlled trials}
We also conducted a simulation study for hybrid-controlled trials.
We assumed that the sample size of the external control data was $N_\text{EC} \in \{50, 100, 200\}$, and the mean outcome of the external control data was set to $\bar y_\text{EC} = 0$.
The null and alternative hypotheses were defined as $\text{H}_0: \delta \ge 0$ and $\text{H}_1: \delta < 0$, respectively, and $\delta_0 = -0.6$.
The known variance was set to $\tau^2 = 1$.
To specify $w$, we calculated the ESS of the analysis prior for the control arm $\pi^{(a)}(\theta_\text{C}\mid \boldsymbol y_\text{EC})$ using the expected local-information-ratio (ELIR) method \citep{neuenschwander-PredictivelyConsistentPrior} implemented in the R package \texttt{RBesT} \citep{weber-ApplyingMetaAnalyticPredictivePriors-2021}.
The rejection criterion for the null hypothesis was defined as the posterior probability of the alternative hypothesis $\Pr(\delta < 0 \mid \boldsymbol{y}_\text{CC}, \boldsymbol{y}_\text{CT}, \boldsymbol{y}_\text{EC}) \ge 0.975$.
The allocation ratio of the sample sizes for the control and treatment arms in the current trial was set to $N_\text{CC} : N_\text{CT} = 1 : 2$.

\subsubsection{Results}
Table \ref{tab: res_sim_two} summarizes the required sample sizes and quantified weights.
The frequentist design without external data required sample sizes of $N_\text{CC} = 33$ and $N_\text{CT} = 66$.
In contrast, the Bayesian design with external control borrowing required smaller sample sizes for all $N_\text{EC}$ values.
Correspondingly, the quantified weight parameter $w$ exhibited a decline as $N_\text{EC}$ increased, reflecting calibration to prevent full borrowing from external control data.

\begin{table}[H]

	\caption{Required sample sizes and quantified weights}
	\centering
	\begin{tabular}{ccccc}
		\hline
		$N_\text{EC}$ & $w$   & $N_\text{CC}$ & $N_\text{CT}$ & $N_\text{CC} + N_\text{CT}$ \\ \hline
		50            & 0.699 & 26            & 52            & 78                          \\
		100           & 0.422 & 26            & 52            & 78                          \\
		200           & 0.252 & 27            & 54            & 81                          \\ \hline
	\end{tabular}
	\label{tab: res_sim_two}
	\vspace{0.7cm}
\end{table}

\section{Case study}\label{sec6}
\subsection{Design}
To illustrate the practical application of the proposed sample size determination method, we conducted a case study.
In this case, we considered a pediatric single-arm clinical trial of litifilimab, a humanized monoclonal antibody targeting blood dendritic cell antigen 2 (BDCA2), for cutaneous lupus erythematosus (CLE), for which external data from an adult clinical trial were available.
CLE is a chronic autoimmune disease that primarily affects the skin, leading to lesions and rashes.
Due to the rarity of CLE in pediatric populations, conducting large-scale clinical trials is challenging in children.
Therefore, leveraging external data from adult trials is highly beneficial when designing such pediatric trials.

We used external data from a Phase II clinical study of litifilimab in adult patients with CLE (LILAC trial) \citep{werth-TrialAntiBDCA2Antibody-2022}.
The LILAC trial was a multi-arm randomized clinical trial comparing litifilimab at doses of 50, 150, and 450 mg with placebo.
For our analysis, we specifically utilized data from the 50 mg dose group as historical evidence to inform the design of the subsequent trial.
The primary endpoint was the percentage change from baseline in the Cutaneous Lupus Erythematosus Disease Area and Severity Index Activity (CLASI-A) score at week 16.
From the LILAC trial data, we set the mean percent change in CLASI-A score at week 16 for the 50 mg litifilimab group as $\bar y_\text{E} = -38.8$ with $\tau = 38.2$ and $N_\text{E} = 26$.
We defined the null and alternative hypotheses as $\text{H}_0: \theta \ge -14.5$ and $\text{H}_1: \theta < -14.5$, respectively, because the mean percent change in CLASI-A in the placebo group was $-14.5$.
The target type I error rate and power were set to $\alpha = 0.025$ and $1-\beta = 0.80$, respectively, and the rejection criterion for the null hypothesis was defined as $T(\boldsymbol{\tilde{y}}_\text{C}) \ge 0.975$.
A grid search for the weight parameter $w$ was performed over the set $\{0, 0.01, \dots, 1\}$.

As in the simulation study for single-arm trials, after determining the sample size $N_\text{C}$ and weight $w$, we evaluated the operating characteristics and estimation performance of the treatment effect by assuming the true treatment effect $\theta \in \{-62.5, \allowbreak -56.5, \allowbreak \dots, 9.5\}$.

\subsection{Results}
The sets of required sample sizes and weights $(N_\text{C}, w)$ determined by the existing and proposed methods were $(21, 0.61)$ and $(17, 0.44)$, respectively, whereas the standard frequentist method required 20 subjects.
Although the existing method borrows external data, its required sample size was essentially the same as that of the frequentist design without borrowing. This is consistent with the simulation results (Table \ref{tab: res_sim}) and indicates that, under the existing NSP, the relaxation of type I error control is not achieved.
Under the proposed NSP, borrowing external data reduced the required sample size relative to the frequentist design.

We then evaluated the operating characteristics and estimation performance of the designs determined above, assuming the true treatment effect $\theta \in \{-62.5, \allowbreak -56.5, \allowbreak \dots, 9.5\}$.
Detailed results, including the rejection rates of the null hypothesis, posterior means with 95\% CIs, and estimation biases, are provided in the Supplementary Materials.

Consistent with the results of the simulation study, the proposed method achieved higher statistical power than that of the existing method over a wider range of the alternative hypothesis, despite requiring a smaller sample size.
However, in the region where the treatment effect is small or worse (i.e., $\theta \ge -14.5$), the rejection rate of the proposed method was higher than that of the existing method.

Regarding the estimation performance, the behavior was similar to that in the simulation.
The proposed method yielded a smaller bias than that of the existing method when the heterogeneity between the external and current trial data was small.
Specifically, in the region where $\theta$ was within $\Theta_0$, the bias of the existing method increased with $\theta$, whereas that of the proposed method approached zero.
However, in the intermediate range ($\theta \in [-38.5, -14.5]$), the existing method showed a slightly smaller bias than that of the proposed method.

\section{Discussion}\label{sec7}
In this study, we proposed an alternative specification of the sampling priors within the Bayesian trial design framework of Psioda and Ibrahim, aimed at achieving the relaxation of type I error control, and extended it to hybrid-controlled trials.
Simulation studies and a real-world case study of CLE demonstrated that the required sample size under the specification of Psioda and Ibrahim was essentially the same as, or even larger than, that of the frequentist design without borrowing.
This indicates that the relaxation is not achieved when the external data strongly support the alternative hypothesis.
Under the proposed NSP, borrowing external data reduced the required sample size relative to the frequentist design.

The reduction in the required sample size results from relaxing type I error control.
Changing the NSP from the restricted posterior distribution to a normal distribution centered at the threshold $\lambda$ implies a more relaxed assumption under the null hypothesis.
Because the Bayesian type I error rate is averaged over the NSP (Equation \eqref{power_function}), controlling it at $\alpha$ allows the frequentist type I error rate to exceed $\alpha$ near the boundary $\lambda$, while it falls below $\alpha$ for parameter values farther from the boundary of hypothesis space.
For example, in Figure \ref{figs: sim_power}, the frequentist type I error rate was $0.041$ at $\theta = \lambda$.
It is important to note that such inflation of the type I error rate is unavoidable if external data are to reduce the sample size because strict control at every null parameter value precludes power gains from borrowing external data \citep{kopp-schneider-PowerGainsUsing-2020}.
The specification of sampling priors should be agreed upon with stakeholders and regulators, considering the possibility that the treatment effect in the current trial lies in the range of the null hypothesis space, as noted by \citet{best-ClassicalTypeError-}.

\begin{figure}[H]
	\centering
	\includegraphics[width = 0.9\linewidth]{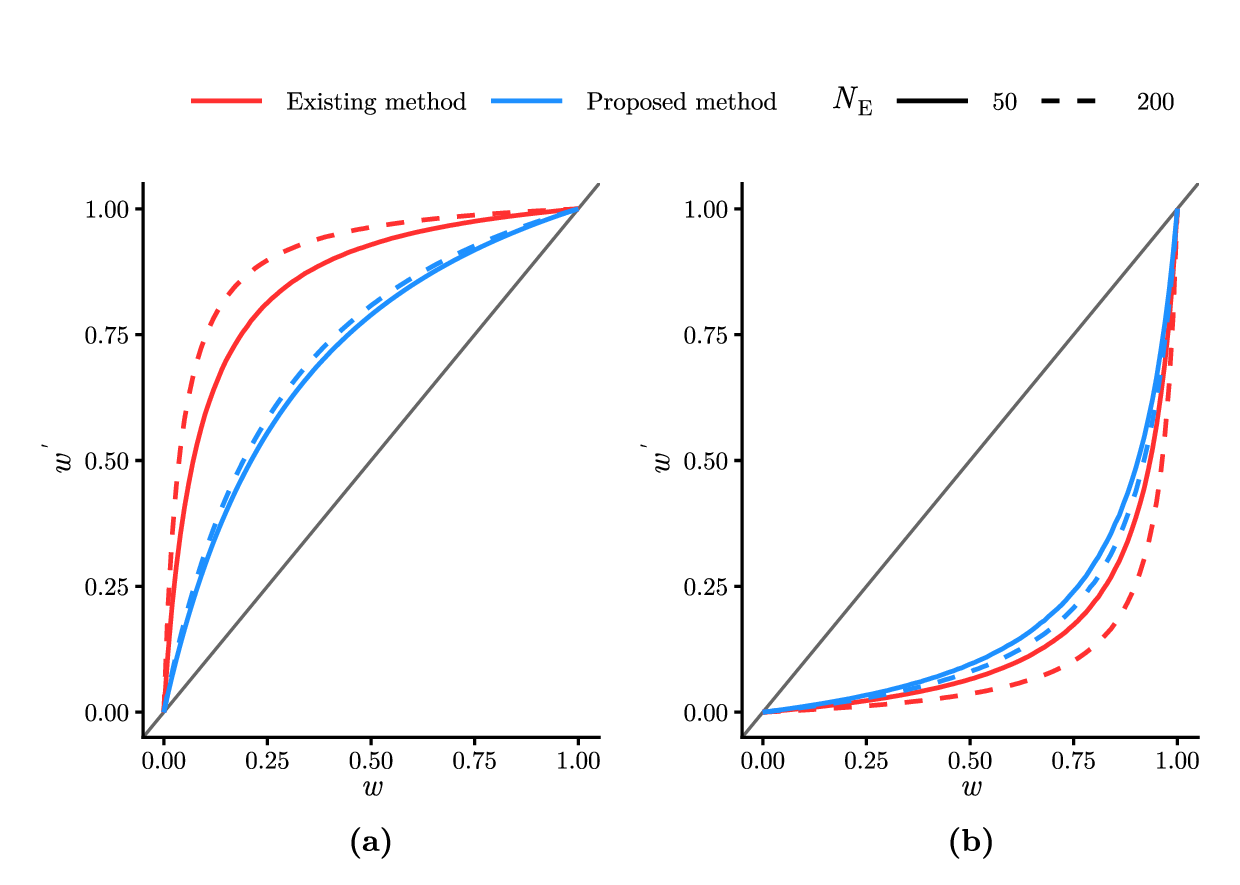}
	\caption{Updated weight $w'$ against the prior weight $w$ for $N_\text{C} = 20$ and $N_\text{E} \in \{50, 200\}$. (a) Under the alternative hypothesis. (b) Under the null hypothesis. The red and blue lines represent the existing and proposed methods, respectively, and the solid and dashed lines correspond to $N_\text{E} = 50$ and $200$, respectively. The gray line represents the identity line $w' = w$.}
	\label{figs: update_w}
\end{figure}

We also observed different behaviors depending on the sample size of the external data $N_\text{E}$ in the single-arm trial scenario.
When $N_\text{E}$ increased from 50 to 200, the proposed method required slightly larger sample sizes, whereas the existing method required smaller sample sizes (Table \ref{tab: res_sim}).
This can be explained by the dependency of the priors on $N_\text{E}$.
Both methods set the variance of the sampling prior to $\tau^2/N_\text{E}$.
As $N_\text{E}$ increases, the probability density around $\theta=\lambda$ (the boundary of the hypothesis) increases under the null sampling prior, complicating the control of the Bayesian type I error.
Furthermore, the specification of the analysis prior plays a crucial role in updating the weight parameter $w$.
With the existing method, the second component has a variance of $\tau^2/N_\text{E}$, implying that both components are highly informative.
This leads to a more rapid update of $w$ (Figure \ref{figs: update_w}); specifically, $w'$ drops to zero more quickly when the data conflict and remains close to one when they are consistent.
By contrast, our proposed method uses a fixed unit-information prior for the second component, independent of $N_\text{E}$.
Consequently, the update of $w$ is more moderate.

Regarding the estimation performance, the proposed method exhibited a smaller bias and less shrinkage than the existing method (Figures \ref{figs: sim_estimate_bias}(a) and \ref{figs: sim_estimate_bias}(b)).
This is a direct result of the analysis prior specification.
The second component of the existing method is as informative as the external data ($N_\text{E}$).
When prior-data conflict occurs ($\theta \ge 0$), the posterior distribution is strongly pulled toward the second component centered at $\lambda$, resulting in significant bias and artificially narrow credible intervals.
On the contrary, the analysis prior with a weakly informative second component prevents such strong shrinkage toward the second component because the analysis prior becomes less informative under prior-data conflict.
Although this leads to slightly wider credible intervals than those of the existing method, it provides a more honest quantification of uncertainty and reduces bias, which is desirable for confirmatory clinical trials.

The Bayesian design for hybrid-controlled trials also demonstrated that it can achieve a sample size reduction relative to the frequentist design (Table \ref{tab: res_sim_two}).
This method specifies the weight parameter $w$ to ensure that the total information in the external and current control data is comparable to that in the treatment data, which prevents the selection of an unrealistically small required sample size that could arise from full borrowing.
Notably, as $N_\text{EC}$ increased, the total sample size required for the current trial also increased.
One possible explanation is that the variance of the sampling prior for the control arm decreases as $N_\text{EC}$ increases, leading to a more rapid update of $w$ (Figure \ref{figs: update_w}).
As shown in \eqref{posterior w calc}, when $N_\text{EC}$ is large, $w$ can be updated to relatively small values even if the heterogeneity between the external control and the current control data is modest.
A smaller $w$ leads to less borrowing, which may result in lower power and a larger required sample size.
Another reason is that the variance of the sampling priors for both the control and treatment arms decreases as $N_\text{EC}$ increases, which may lead to a smaller variance of $\delta$ and a more conservative design.
Future research should investigate the performance of the proposed method in hybrid-controlled trials under various scenarios of heterogeneity between the external control and current trial data and explore potential modifications to the method to better handle such heterogeneity.

Although this study assumed the availability of a single external data source, borrowing from multiple external data sources may be desirable in practice.
The proposed framework can be naturally applied to such a scenario.
Potential extensions that accommodate multiple external datasets include methods such as a meta-analytic-predictive (MAP) prior \citep{schmidli-RobustMetaanalyticpredictivePriors-2014}, a power prior \citep{banbeta-ModifiedPowerPrior-2019}, a potential bias model with shrinkage prior \citep{ohigashi-UsingHorseshoePrior-2022, ohigashiPotentialBiasModels2025}, and a nonparametric Bayesian method that clusters external and current controls according to similarity in the parameters of interest \citep{ohigashi-NonparametricBayesianApproach-2025}.
Notably, if a MAP prior is approximated by a normal distribution, the conjugacy properties are preserved, allowing for efficient analytical posterior calculations, similar to our current approach.

Our method has a few limitations.
As described in Section \ref{sec3}, the second component of the analysis prior is specified as a unit-information prior.
While this specification is commonly used in robust mixture priors \citep{schmidli-RobustMetaanalyticpredictivePriors-2014, Weru05052026}, its relative informativeness may vary depending on the sample size of the external data.
Therefore, the quantified $w$ identified by the proposed procedure may depend on the specification of the second component.
Recent work by \citet{ratta-InterplayPriorWeight-2026} has emphasized the importance of jointly determining not only $w$, but also the mean and variance of the second component in a robust mixture prior.
The objective of this study is to propose and extend the clinical trial design framework to determine the required sample size, rather than to optimize the design parameters of the analysis prior.
Further research should explore the selection of these parameters, without being limited to $w$, during the trial design stage.

In addition, we assumed a known variance for the outcome.
In practice, the variance is often unknown and must be estimated from the data.
The framework can be extended to the case of unknown variance.
For example, with a normal-inverse-gamma prior, each component of the analysis prior remains conjugate and the posterior distribution can be derived analytically.

We focused on continuous endpoints with a normal distribution.
This framework can be easily extended to binary outcomes using a conjugate beta prior or by taking the log-odds ratio for binary data.
However, applying this method to time-to-event outcomes (survival analysis) is likely to result in a loss of conjugacy.
In such non-conjugate settings, the posterior distributions cannot be derived analytically, necessitating computationally intensive methods such as Markov Chain Monte Carlo (MCMC) within the simulation procedure for sample size determination.

In this study, the degree of borrowing from external data was adjusted depending on the heterogeneity of outcomes between the external and current data, without considering the potential sources of heterogeneity, such as covariates.
When the external population differs from the current trial population due to covariates such as age or comorbidities, simply borrowing the marginal distribution may introduce bias.
Future research should investigate incorporating covariate adjustments into prior construction to better account for the exchangeability between populations and further validate the borrowed information.
Propensity score methods \citep{lin-PropensityscorebasedPriorsBayesian-2019, wang-PropensityScoreintegratedComposite-2020, liu-PropensityscorebasedMetaanalyticPredictive-2021} or the latent exchangeability prior (LEAP) \citep{alt-LEAPLatentExchangeability-2024} can be useful for this purpose.

In conclusion, we proposed an alternative specification of the null sampling prior to enable the relaxation of type I error control described by Psioda and Ibrahim even when the external data strongly support the alternative hypothesis, thereby allowing external data to reduce the required sample size.
This reduction is achieved at the cost of a frequentist type I error rate exceeding the nominal level near the null boundary, and the degree of relaxation should be agreed upon with stakeholders and regulators.
Furthermore, the extension to two-arm trials with external control data provides a practical design option when external data are available only for the control arm.
The proposed framework can improve the feasibility of clinical trials in rare diseases and pediatric populations, where recruiting a large number of patients is challenging.

\section*{Financial disclosure}

This work was supported by Japan Society for the Promotion of Science KAKENHI (Grant Number 24K02662).

\section*{Conflict of interest}

The authors declare no potential conflicts of interest.

\section*{Supporting Information}

Simulation and case study results for all scenarios are provided in the Supplementary Materials.
Simulation and case study were performed using R version 4.5.2 (R Foundation for Statistical Computing, Vienna, Austria)\citep{R-teams} and R code is available at \url{https://github.com/W-Murasaki/Modification-and-Extension-of-Bayesian-Clinical-Trial-Design-using-external-data}.



\clearpage
\appendix
\section{Equivalence between Bayesian and Frequentist Type I Error Rates with Point Mass Sampling Prior and Non-informative Analysis Prior} \label{app: bayesian_freq_equivalence}

Consider the one-sided hypothesis test:
\begin{align*}
	\text{H}_0: \theta \in \Theta_0 = \{\theta : \theta \ge \lambda\} \quad \text{versus} \quad \text{H}_1: \theta \in \Theta_1 = \{\theta : \theta < \lambda\},
\end{align*}
where $\lambda$ is the clinically meaningful threshold. Assume the observed outcomes $y_{\text{C}, i_\text{C}}$ from the current trial follow a normal distribution with a known variance $\tau^2$:
\begin{align*}
	y_{\text{C}, i_\text{C}} \overset{\text{i.i.d.}}{\sim} N(\theta, \tau^2) \quad (i_\text{C} = 1,\ldots,N_\text{C}).
\end{align*}

The rejection criterion is defined as: reject $\text{H}_0$ if $\Pr(\theta \in \Theta_1 \mid \boldsymbol{y}_\text{C}, \boldsymbol{y}_\text{E}) \ge \phi$, where $\phi$ is the posterior probability threshold (e.g., $\phi = 0.975$ for $\alpha = 0.025$). When the analysis prior is non-informative, the posterior distribution is
\begin{align*}
	\pi(\theta \mid \boldsymbol{y}_\text{C}, \boldsymbol{y}_\text{E}) = f_N\left(\theta \mid \bar{y}_\text{C}, \frac{\tau^2}{N_\text{C}}\right).
\end{align*}

The rejection condition becomes:
\begin{align}
	\Pr(\theta < \lambda \mid \boldsymbol{y}_\text{C}, \boldsymbol{y}_\text{E})                                                                                                                    & \ge \phi \nonumber \\
	\Pr\left(\frac{\theta - \bar{y}_\text{C}}{\sqrt{\tau^2/N_\text{C}}} < \frac{\lambda - \bar{y}_\text{C}}{\sqrt{\tau^2/N_\text{C}}} \mid \boldsymbol{y}_\text{C}, \boldsymbol{y}_\text{E}\right) & \ge \phi \nonumber \\
	\Pr\left(Z < \frac{\lambda - \bar{y}_\text{C}}{\sqrt{\tau^2/N_\text{C}}}\right)                                                                                                                & \ge \phi \nonumber \\
	\frac{\lambda - \bar{y}_\text{C}}{\sqrt{\tau^2/N_\text{C}}}                                                                                                                                    & \ge z_{\phi},
\end{align}
where $Z \sim N(0,1)$ is the standard normal distribution and $z_{\phi}$ is the $\phi$-quantile of the standard normal distribution (e.g., $z_{0.975} = 1.96$). This can be rewritten as $\bar{y}_\text{C} \le \lambda - z_{\phi}\sqrt{\tau^2/N_\text{C}}$.

The Bayesian type I error rate under the null hypothesis $\text{H}_0$ is defined as:
\begin{align}
	\alpha_\text{Bayes} & = \mathrm{E}_{\pi^{(s)}_0}[\mathds{1}\{\Pr(\theta \in \Theta_1 \mid \boldsymbol{y}_\text{C}, \boldsymbol{y}_\text{E}) \ge \phi\}] \nonumber                                                                                              \\
	                    & = \int \mathds{1}\{\Pr(\theta \in \Theta_1 \mid \boldsymbol{y}_\text{C}, \boldsymbol{y}_\text{E}) \ge \phi\} \left(\int f_N(\boldsymbol{y}_\text{C} \mid \theta, \tau^2) \pi^{(s)}_0(\theta) \, d\theta\right) d\boldsymbol{y}_\text{C},
\end{align}
where $\pi^{(s)}_0(\theta)$ is the sampling prior under the null hypothesis. When the sampling prior is a point mass at $\lambda$ (i.e., $\pi^{(s)}_0(\theta) = \delta(\theta - \lambda)$), the type I error rate reduces to:
\begin{align}
	\alpha_\text{Bayes} = \int \mathds{1}\{\bar{y}_\text{C} \le \lambda - z_{\phi}\sqrt{\tau^2/N_\text{C}}\} f_N(\boldsymbol{y}_\text{C} \mid \lambda, \tau^2) d\boldsymbol{y}_\text{C}.
\end{align}

Since $\bar{y}_\text{C} \mid \theta = \lambda \sim N(\lambda, \tau^2/N_\text{C})$, we have:
\begin{align}
	\alpha_\text{Bayes} & = \Pr(\bar{y}_\text{C} \le \lambda - z_{\phi}\sqrt{\tau^2/N_\text{C}} \mid \theta = \lambda) \nonumber \\
	                    & = \Pr\left(\frac{\bar{y}_\text{C} - \lambda}{\sqrt{\tau^2/N_\text{C}}} \le -z_{\phi}\right) \nonumber  \\
	                    & = \Pr(Z \le -z_{\phi}) \nonumber                                                                       \\
	                    & = \Phi(-z_{\phi}) \nonumber                                                                            \\
	                    & = 1 - \Phi(z_{\phi}),
\end{align}
where $\Phi$ is the cumulative distribution function of the standard normal distribution. For the standard frequentist test, the type I error rate when the true parameter is $\theta = \lambda$ is $\alpha_\text{Freq} = 1 - \Phi(z_{\phi})$. Therefore, we have demonstrated that $\alpha_\text{Bayes} = \alpha_\text{Freq} = 1 - \Phi(z_{\phi})$. This shows that when a point mass sampling prior is combined with a non-informative analysis prior, the Bayesian type I error rate coincides with the frequentist type I error rate evaluated at the null boundary $\theta = \lambda$.

\section{Derivation of Posterior Weight} \label{app: calc_posterior_ap_and_w}

Let the analysis prior $\pi^{(a)}(\theta\mid \boldsymbol y_\text{E})$ be
\begin{align*}
	\pi^{(a)}(\theta\mid \boldsymbol y_\text{E}) & = wf_N(\theta \mid \mu_1, \sigma_1^2) + (1-w)f_N(\theta \mid \mu_2, \sigma_2^2) \\
	                                             & = wf_1(\theta) + (1-w)f_2(\theta)
\end{align*}

Assuming that the external and current data are independent, the posterior distribution $\pi(\theta\mid \boldsymbol y_\text{C}, \boldsymbol y_\text{E})$ is calculated by Bayes' theorem as follows:
\begin{equation} \label{app_2: post_theta}
	\begin{split}
		\pi(\theta\mid \boldsymbol y_\text{C}, \boldsymbol y_\text{E}) & = \frac{1}{m(\boldsymbol y_\text{C}\mid \boldsymbol y_\text{E})}\pi^{(a)}(\theta\mid \boldsymbol y_\text{E})L(\theta \mid \boldsymbol y_\text{C}, \boldsymbol y_\text{E})                 \\
		                                                               & = \frac{1}{m(\boldsymbol y_\text{C}\mid \boldsymbol y_\text{E})}wf_1(\theta)L(\theta \mid \boldsymbol y_\text{C})                                                                         \\
		                                                               & \quad +\frac{1}{m(\boldsymbol y_\text{C}\mid \boldsymbol y_\text{E})}(1-w)f_2(\theta)L(\theta \mid \boldsymbol y_\text{C})                                                                \\
		                                                               & = \frac{m_1(\boldsymbol y_\text{C})}{m(\boldsymbol y_\text{C}\mid \boldsymbol y_\text{E})}w \frac{f_1(\theta)L(\theta \mid \boldsymbol y_\text{C})}{m_1(\boldsymbol y_\text{C})}          \\
		                                                               & \quad +\frac{m_2(\boldsymbol y_\text{C})}{m(\boldsymbol y_\text{C}\mid \boldsymbol y_\text{E})}(1-w) \frac{f_2(\theta)L(\theta \mid \boldsymbol y_\text{C})}{m_2(\boldsymbol y_\text{C})} \\
		                                                               & =\frac{m_1(\boldsymbol y_\text{C})}{m(\boldsymbol y_\text{C}\mid \boldsymbol y_\text{E})}wf_1(\theta\mid \boldsymbol y_\text{C})                                                          \\
		                                                               & \quad +\frac{m_2(\boldsymbol y_\text{C})}{m(\boldsymbol y_\text{C}\mid \boldsymbol y_\text{E})}(1-w) f_2(\theta\mid \boldsymbol y_\text{C}).
	\end{split}
\end{equation}
Here, $m_1(\boldsymbol y_\text{C})$, $m_2(\boldsymbol y_\text{C})$, and $m(\boldsymbol y_\text{C}\mid \boldsymbol y_\text{E})$ are the marginal likelihoods corresponding to the probability density functions $f_1(\theta)$, $f_2(\theta)$, and $\pi^{(a)}(\theta\mid \boldsymbol y_\text{E})$, respectively.
\begin{align}
	m_1(\boldsymbol y_\text{C})                          & = \int L(\theta \mid \boldsymbol y_\text{C})f_1(\theta) \, d\theta,                                                 \label{app2 m1} \\
	m_2(\boldsymbol y_\text{C})                          & = \int L(\theta \mid \boldsymbol y_\text{C})f_2(\theta) \, d\theta,                                                 \label{app2 m2} \\
	m(\boldsymbol y_\text{C}\mid \boldsymbol y_\text{E}) & = \int L(\theta \mid \boldsymbol y_\text{C})\pi^{(a)}(\theta\mid \boldsymbol y_\text{E}) \, d\theta. \notag
\end{align}

Since the posterior distribution $\pi(\theta\mid \boldsymbol y_\text{C}, \boldsymbol y_\text{E})$ is a probability density function,
\begin{align}\label{app2 1}
	\int \pi(\theta\mid \boldsymbol y_\text{C}, \boldsymbol y_\text{E})\, d\theta = 1.
\end{align}

Furthermore, since $f_1(\theta\mid \boldsymbol y_\text{C})$ and $f_2(\theta\mid \boldsymbol y_\text{C})$ are also probability density functions with respect to $\theta$,
\begin{align}
	\int f_1(\theta\mid \boldsymbol y_\text{C}) \, d\theta & = 1,                \\
	\int f_2(\theta\mid \boldsymbol y_\text{C}) \, d\theta & = 1. \label{app2 3}
\end{align}

From Equations \eqref{app_2: post_theta}, \eqref{app2 1}, and \eqref{app2 3},
\begin{align*}
	\frac{m_1(\boldsymbol y_\text{C})}{m(\boldsymbol y_\text{C}\mid \boldsymbol y_\text{E})}w+\frac{m_2(\boldsymbol y_\text{C})}{m(\boldsymbol y_\text{C}\mid \boldsymbol y_\text{E})}(1-w) =1
\end{align*}

Therefore,
\begin{align}\label{eq1}
	m(\boldsymbol y_\text{C}\mid \boldsymbol y_\text{E}) = m_1(\boldsymbol y_\text{C})w+m_2(\boldsymbol y_\text{C})(1-w).
\end{align}

Substituting Equation \eqref{eq1} into Equation \eqref{app_2: post_theta} and rearranging,
\begin{align*}
	\pi(\theta\mid \boldsymbol y_\text{C}, \boldsymbol y_\text{E}) & =\frac{wm_1(\boldsymbol y_\text{C})}{wm_1(\boldsymbol y_\text{C})+(1-w)m_2(\boldsymbol y_\text{C})} \times f_1(\theta\mid \boldsymbol y_\text{C})            \\
	                                                               & \quad +\frac{(1-w)m_2(\boldsymbol y_\text{C})}{wm_1(\boldsymbol y_\text{C})+(1-w)m_2(\boldsymbol y_\text{C})} \times f_2(\theta\mid \boldsymbol y_\text{C}).
\end{align*}

From the above derivation,
\begin{align}\label{calc post w}
	w' = \frac{wm_1(\boldsymbol y_\text{C})}{wm_1(\boldsymbol y_\text{C})+(1-w)m_2(\boldsymbol y_\text{C})}.
\end{align}

The marginal likelihoods $m_i (\boldsymbol y_\text{C}), \, (i \in \{1,2\})$ can be analytically derived when a normal distribution is assumed. Under the assumption that the current data follow the distribution as in Equation \eqref{dist_new_data},
$$
	L(\theta \mid \boldsymbol y_\text{C}) = \prod_{i_\text{C}=1}^{N_\text{C}} \frac{1}{\sqrt{2\pi\tau^2}}\exp\left\{-\frac{(y_{\text{C},i_\text{C}}-\theta)^2}{2\tau^2}\right\}
$$

Since $f_i(\theta) = f_N(\theta \mid \mu_i, \sigma_i^2)$, Equations \eqref{app2 m1} and \eqref{app2 m2} become
\begin{align*}
	m_i(\boldsymbol y_\text{C}) & = \int L(\theta \mid \boldsymbol y_\text{C})f_i(\theta) \, d\theta                                                                         \\
	                            & = \int \prod_{i_\text{C}=1}^{N_\text{C}} \frac{1}{\sqrt{2\pi\tau^2}}\exp\left\{-\frac{(y_{\text{C},i_\text{C}}-\theta)^2}{2\tau^2}\right\} \\
	                            & \quad \times \frac{1}{\sqrt{2\pi\sigma_i^2}}\exp\left\{-\frac{(\theta - \mu_i)^2}{2\sigma_i^2}\right\} \, d\theta,
\end{align*}

which simplifies to
\begin{align*}
	m_i(\boldsymbol y_\text{C}) = \frac{1}{\sqrt{2\pi\left(\sigma_i^2 + \frac{\tau^2}{N_\text{C}}\right)}}\exp\left\{-\frac{(\bar y_\text{C} - \mu_i)^2}{2\left(\sigma_i^2 + \frac{\tau^2}{N_\text{C}}\right)}\right\}.
\end{align*}

Let $\varphi(\bar y_\text{C} \mid \mu, \sigma^2)$ denote the probability density at $\bar y_\text{C}$ for a normal distribution with mean $\mu$ and variance $\sigma^2$. Then Equation \eqref{calc post w} becomes
\begin{align}\label{post W by m}
	w' = \frac{w\varphi\left(\bar y_\text{C} \mid \mu_1, \sigma_1^2 + \frac{\tau^2}{N_\text{C}}\right)}{w\varphi\left(\bar y_\text{C} \mid \mu_1, \sigma_1^2 + \frac{\tau^2}{N_\text{C}}\right)+(1-w)\varphi\left(\bar y_\text{C} \mid \mu_2, \sigma_2^2 + \frac{\tau^2}{N_\text{C}}\right)}.
\end{align}

\section{Update of the Weight When the Second Component is a Non-informative Prior} \label{app: no_info}

In a normal distribution, as the variance increases, the distribution becomes flatter and approaches a uniform distribution. Therefore, to make the second component $f_N(\theta \mid \mu_2, \sigma_2^2)$ of the analysis prior a non-informative prior, we set $\sigma_2 \rightarrow \infty$.

In this case, the probability density $\varphi(\bar y_\text{C} \mid \mu_2, \sigma_2^2 + \tau^2/N_\text{C})$ becomes
\begin{align*}
	\varphi\left(\bar y_\text{C} \mid \mu_2, \sigma_2^2 + \frac{\tau^2}{N_\text{C}}\right) \overset{\sigma_2 \rightarrow \infty}{\longrightarrow} 0.
\end{align*}

Therefore, unless $w$ is extremely small, $w'$ from Equation \eqref{post W by m} approaches
\begin{align*}
	w' & = \frac{w\varphi\left(\bar y_\text{C} \mid \mu_1, \sigma_1^2 + \frac{\tau^2}{N_\text{C}}\right)}{w\varphi\left(\bar y_\text{C} \mid \mu_1, \sigma_1^2 + \frac{\tau^2}{N_\text{C}}\right)+(1-w)\varphi\left(\bar y_\text{C} \mid \mu_2, \sigma_2^2 + \frac{\tau^2}{N_\text{C}}\right)} \\
	   & \overset{\sigma_2 \rightarrow \infty}{\longrightarrow} \frac{w\varphi\left(\bar y_\text{C} \mid \mu_1, \sigma_1^2 + \frac{\tau^2}{N_\text{C}}\right)}{w\varphi\left(\bar y_\text{C} \mid \mu_1, \sigma_1^2 + \frac{\tau^2}{N_\text{C}}\right)} = 1.
\end{align*}

This means that regardless of the magnitude of heterogeneity in treatment effects between external and current data, $w'$ always approaches $1$. Consequently, it becomes difficult to dynamically adjust the degree of borrowing from external data according to treatment effect heterogeneity.

To confirm these findings numerically, we compared the results of $w'$ and $\alpha$ when the second component of the analysis prior is a non-informative prior versus the proposed analysis prior for $w \in \{0, 0.1, \ldots, 1\}$. Table \ref{tab: comp_no_info} presents the results of $w'$ and $\alpha$ for the scenario with $N_\text{C} = 20$, the true value of the treatment effect $\theta = 0$, external data sample size $N_\text{E} = 50$, and mean of external data $\bar y_\text{E} = -0.6$.

From Table \ref{tab: comp_no_info}, when a non-informative prior is used as the second component, $w'$ tends to approach $1$, and the degree of borrowing from external data increases even when treatment effect heterogeneity is substantial. As a result, $\alpha$ rapidly increases.

\begin{table}[htbp]
	\centering
	\caption{Results of $w'$ and $\alpha$ for $w \in \{0, 0.1, \ldots, 1\}$ when $N_\text{E} = 50$, $\bar y_\text{E} = -0.6$, and $N_\text{C} = 20$. Non-informative: analysis prior with non-informative second component; Unit-information: analysis prior with unit-information second component.}
	\begin{tabular}{ccccc} \hline \label{tab: comp_no_info}
		                      & \multicolumn{2}{c}{Non-informative} & \multicolumn{2}{c}{Unit-information}                     \\ \cline{2-5}
		\multirow{-2}{*}{$w$} & $w'$                                & $\alpha$                             & $w'$   & $\alpha$ \\ \hline

		0                     & 0                                   & 0.0248                               & 0      & 0.0220   \\
		0.1                   & 0.7617                              & 0.2609                               & 0.0317 & 0.0307   \\
		0.2                   & 0.8785                              & 0.3628                               & 0.0690 & 0.0405   \\
		0.3                   & 0.9253                              & 0.4377                               & 0.1127 & 0.0507   \\
		0.4                   & 0.9508                              & 0.5015                               & 0.1652 & 0.0626   \\
		0.5                   & 0.9664                              & 0.5583                               & 0.2277 & 0.0774   \\
		0.6                   & 0.9774                              & 0.6147                               & 0.3069 & 0.0957   \\
		0.7                   & 0.9854                              & 0.6741                               & 0.4081 & 0.1219   \\
		0.8                   & 0.9914                              & 0.7390                               & 0.5430 & 0.1619   \\
		0.9                   & 0.9962                              & 0.8211                               & 0.7273 & 0.2393   \\
		1                     & 1                                   & 0.9988                               & 1      & 0.9988   \\ \hline
	\end{tabular}
\end{table}

\clearpage
\thispagestyle{empty}
\vspace*{\fill}

\begin{center}
	{\Huge\bfseries Supplementary Materials} \\[1.5cm]

\end{center}

\vspace*{\fill}

\clearpage

\setcounter{figure}{0}
\renewcommand{\thefigure}{S\arabic{figure}}
\setcounter{table}{0}
\renewcommand{\thetable}{S\arabic{table}}

\setcounter{section}{0}
\renewcommand{\thesection}{S\arabic{section}}
\renewcommand{\thesubsection}{\thesection.\arabic{subsection}}

\section{Power, Posterior Mean, and Bias for Case Study}

\begin{figure}[H]
	\centering
	\includegraphics[width = 0.9\linewidth]{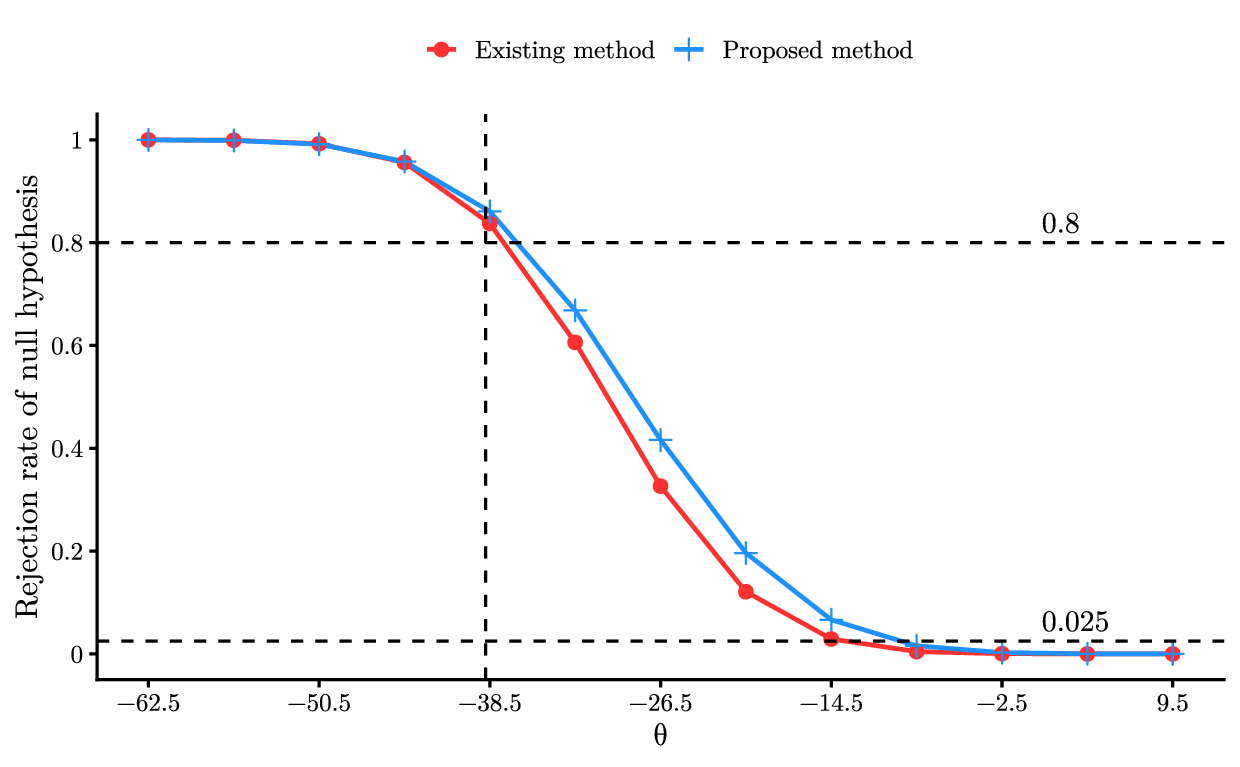}
	\caption{Rejection rates of the null hypothesis in the real data analysis. The sample sizes $N_\mathrm{C}$ were $21$ and $17$ for the existing and proposed methods, respectively.}
	\label{figs: app_power}
\end{figure}

\begin{figure}[H]
	\centering
	\includegraphics[width = 0.9\linewidth]{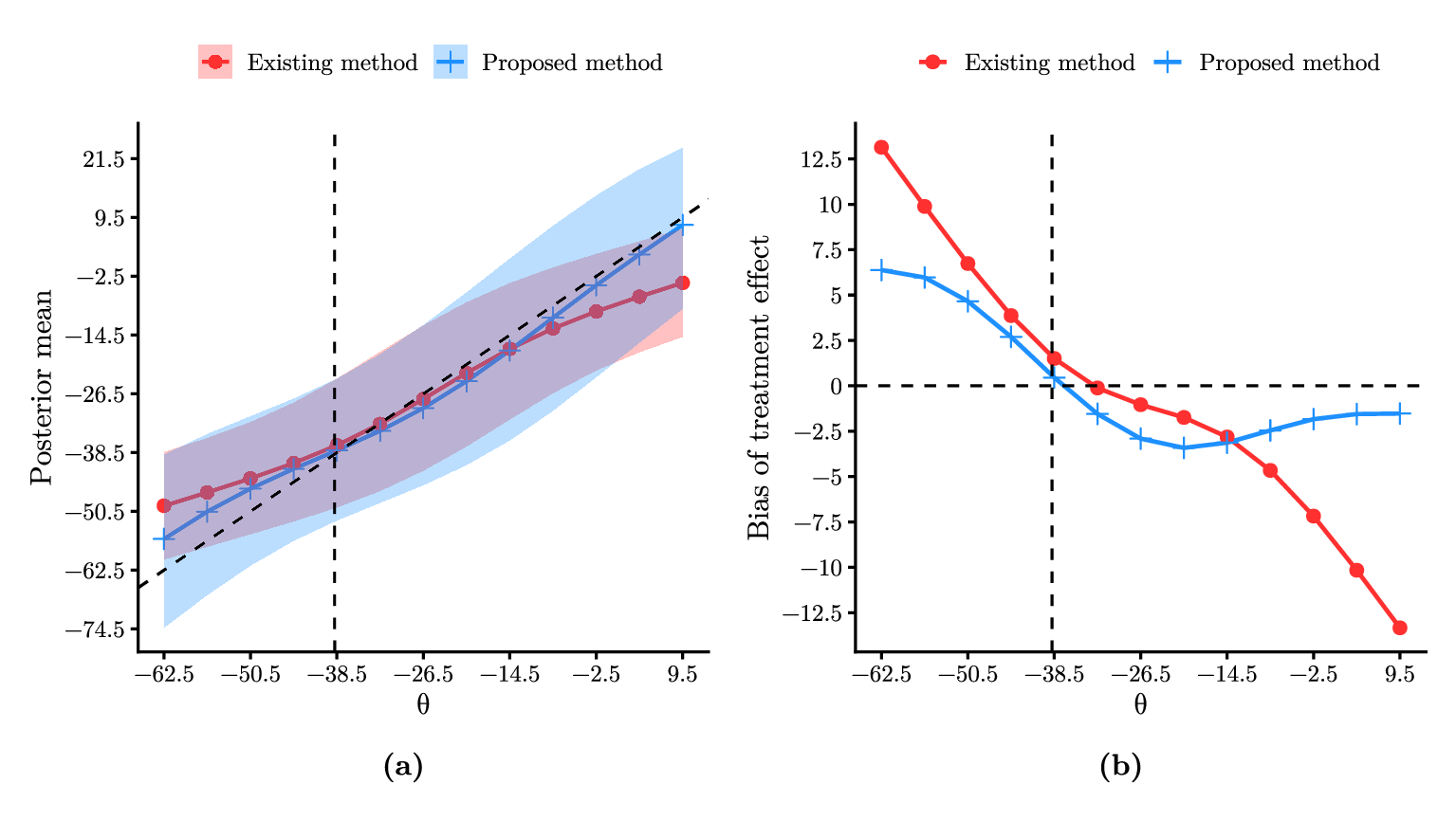}
	\caption{Posterior means, 95\% credible intervals (95\% CIs) and biases of the treatment effect estimates in the real data analysis. The sample sizes $N_\mathrm{C}$ for the existing and proposed methods were $21$ and $17$, respectively.}
	\label{figs: app_posterior_and_bias}
\end{figure}

\section{Power, Posterior Mean, and Bias for Other Scenarios}
\subsection{$\bar y_\mathrm{E} = -0.4$ and $N_\mathrm{E} = 50$}
\begin{figure}[H]
	\centering
	\includegraphics[width = 0.9\linewidth]{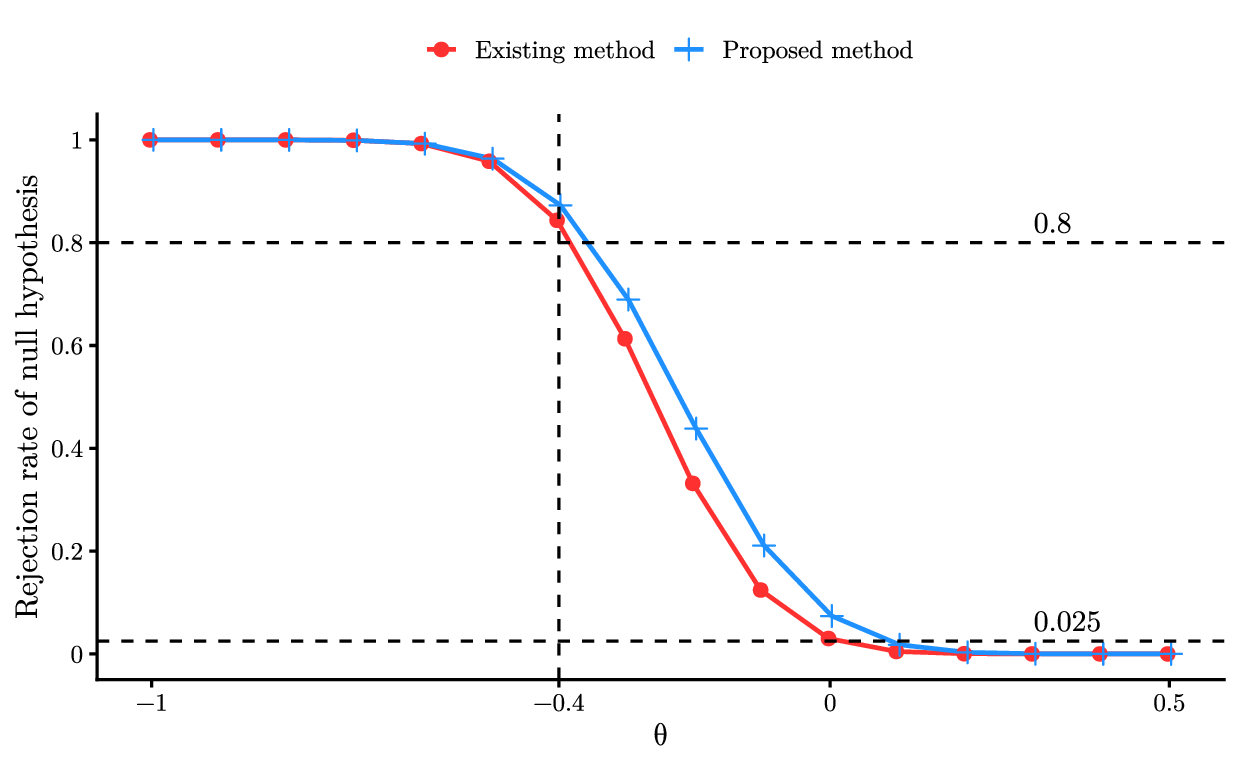}
	\caption{Rejection rates of the null hypothesis for $\bar y_\mathrm{E} = -0.4$ and $N_\mathrm{E} = 50$. The sample sizes $N_\mathrm{C}$ for the existing and proposed methods were 52 and 42, respectively.}
\end{figure}

\begin{figure}[H]
	\centering
	\includegraphics[width = 0.9\linewidth]{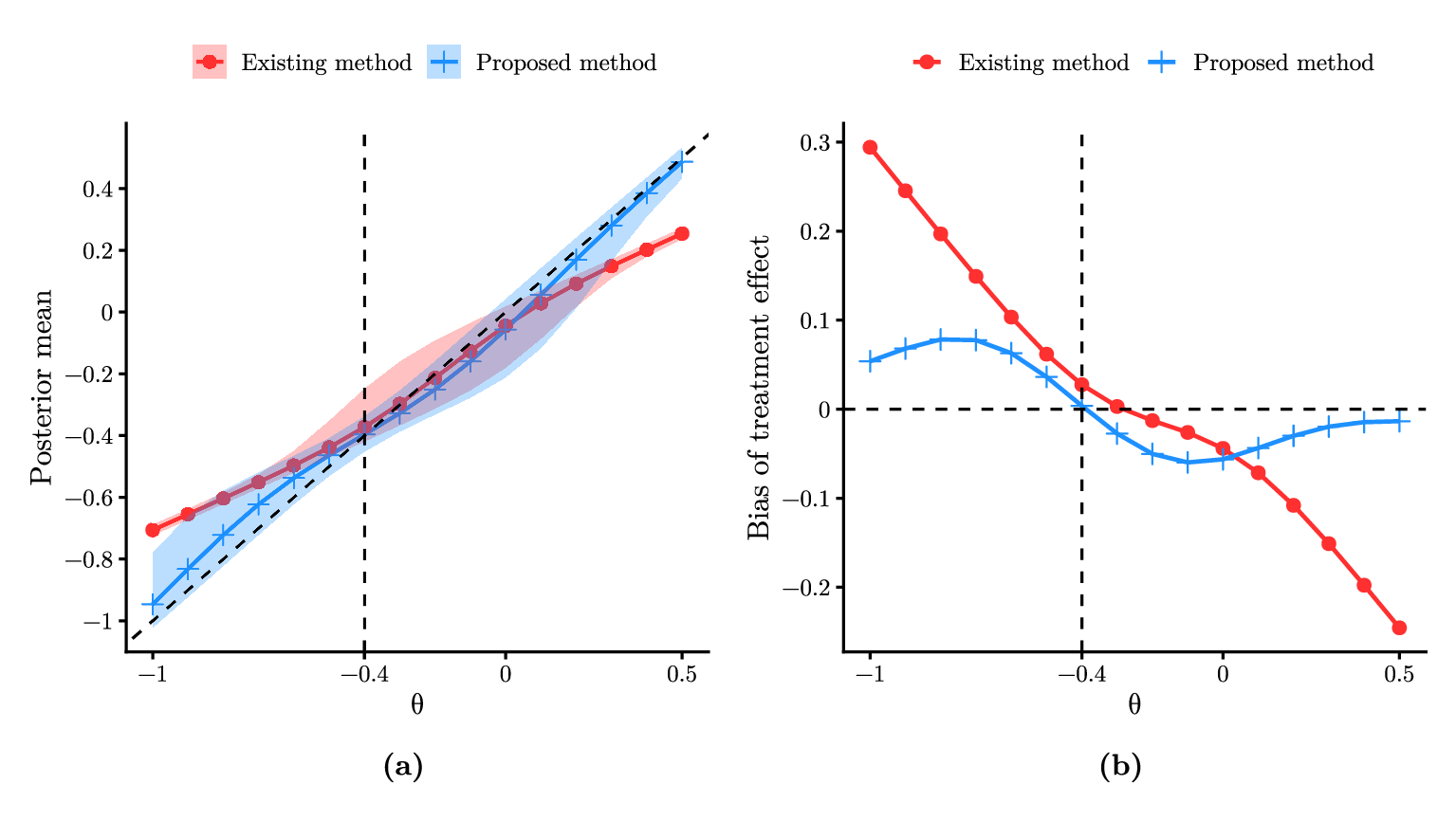}
	\caption{Posterior means, 95\% credible intervals (95\% CIs) and biases of the treatment effect estimates for $\bar y_\mathrm{E} = -0.4$ and $N_\mathrm{E} = 50$. The sample sizes $N_\mathrm{C}$ for the existing and proposed methods were 52 and 42, respectively.}
\end{figure}

\subsection{$\bar y_\mathrm{E} = -0.6$ and $N_\mathrm{E} = 50$}
\begin{figure}[H]
	\centering
	\includegraphics[width = 0.9\linewidth]{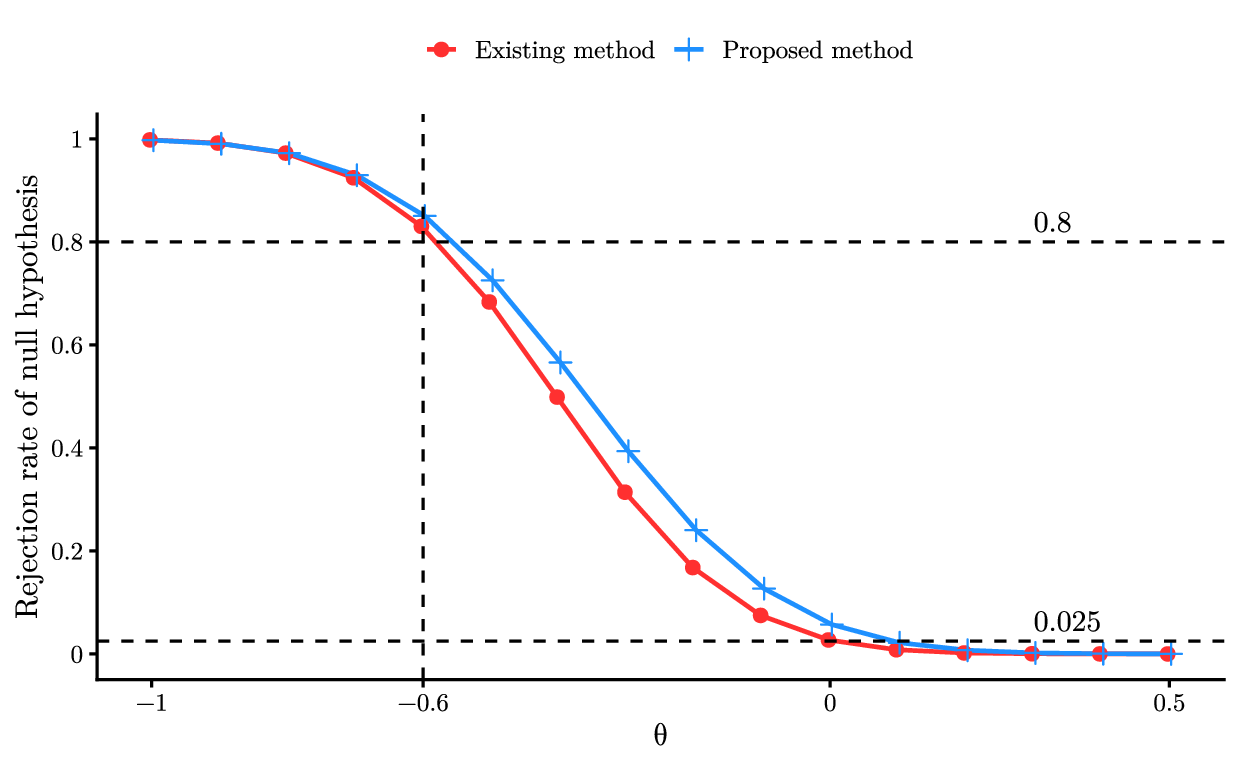}
	\caption{Rejection rates of the null hypothesis for $\bar y_\mathrm{E} = -0.6$ and $N_\mathrm{E} = 50$. The sample sizes $N_\mathrm{C}$ for the existing and proposed methods were 23 and 19, respectively.}
\end{figure}

\begin{figure}[H]
	\centering
	\includegraphics[width = 0.9\linewidth]{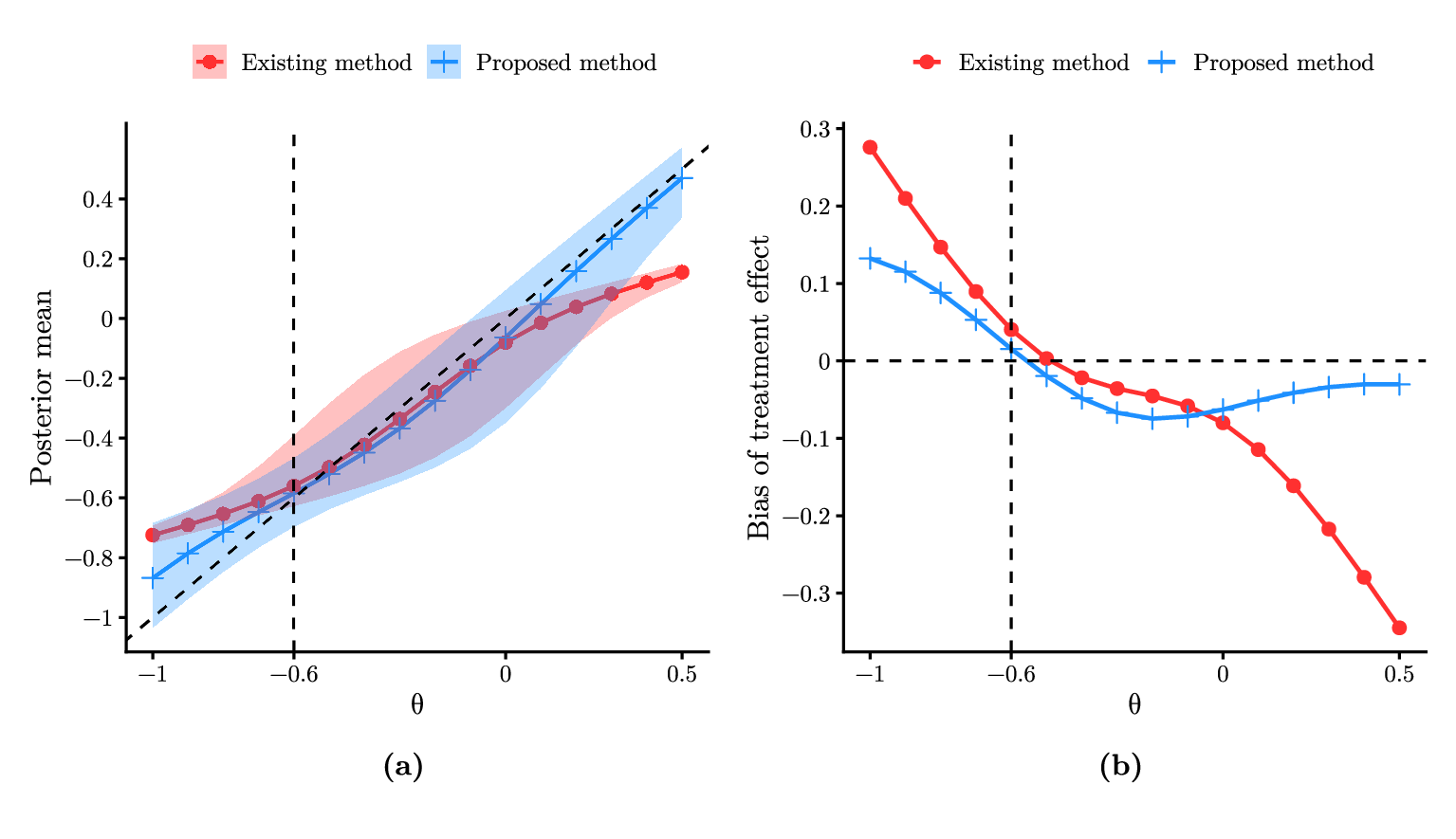}
	\caption{Posterior means, 95\% credible intervals (95\% CIs) and biases of the treatment effect estimates for $\bar y_\mathrm{E} = -0.6$ and $N_\mathrm{E} = 50$. The sample sizes $N_\mathrm{C}$ for the existing and proposed methods were 23 and 19, respectively.}
\end{figure}

\subsection{$\bar y_\mathrm{E} = -0.8$ and $N_\mathrm{E} = 50$}
\begin{figure}[H]
	\centering
	\includegraphics[width = 0.9\linewidth]{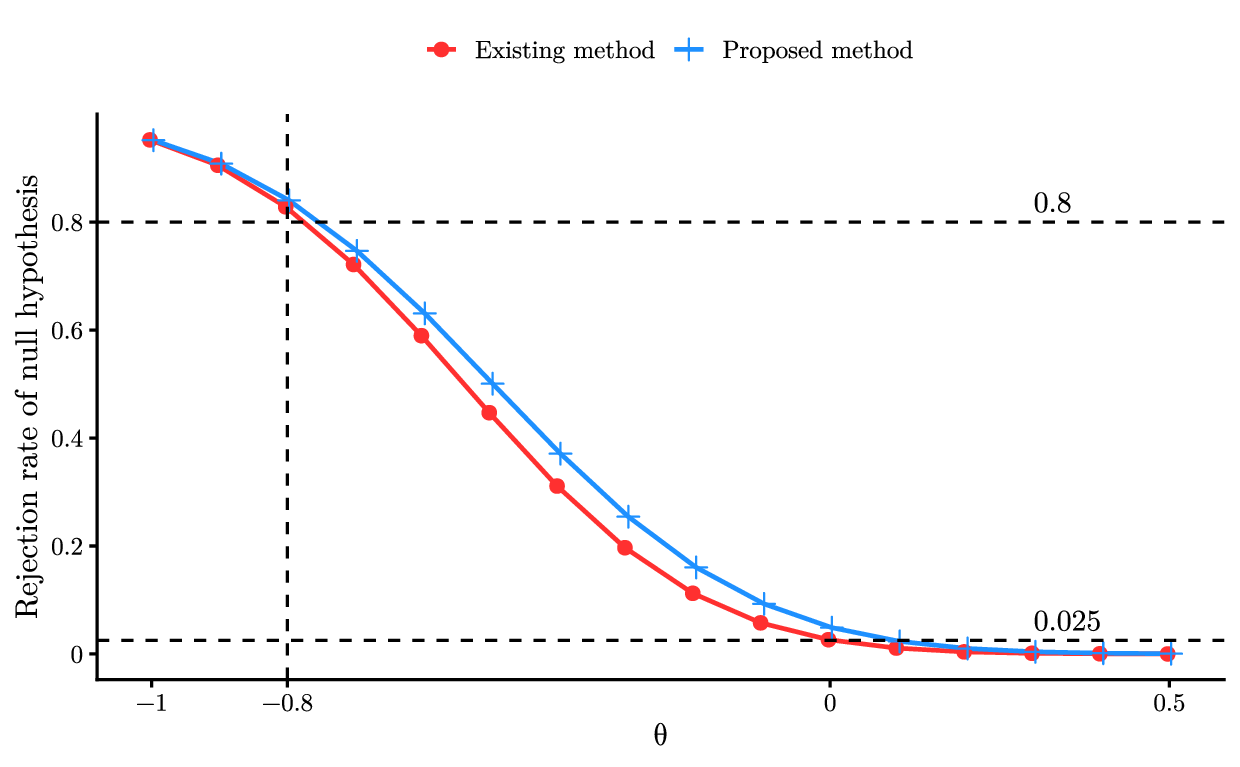}
	\caption{Rejection rates of the null hypothesis for $\bar y_\mathrm{E} = -0.8$ and $N_\mathrm{E} = 50$. The sample sizes $N_\mathrm{C}$ for the existing and proposed methods were 13 and 11, respectively.}
\end{figure}

\begin{figure}[H]
	\centering
	\includegraphics[width = 0.9\linewidth]{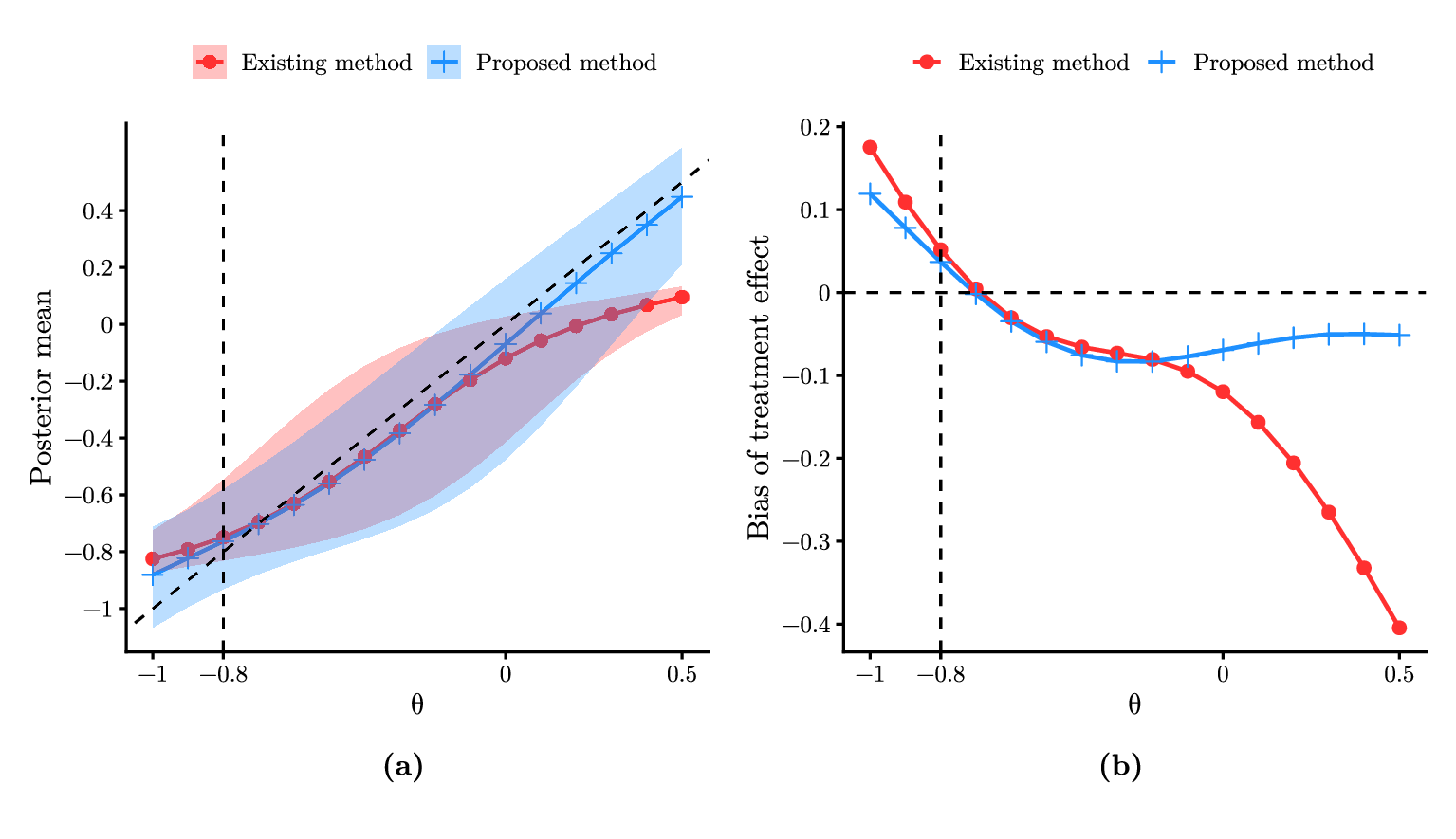}
	\caption{Posterior means, 95\% credible intervals (95\% CIs) and biases of the treatment effect estimates for $\bar y_\mathrm{E} = -0.8$ and $N_\mathrm{E} = 50$. The sample sizes $N_\mathrm{C}$ for the existing and proposed methods were 13 and 11, respectively.}
\end{figure}

\subsection{$\bar y_\mathrm{E} = -1.0$ and $N_\mathrm{E} = 50$}
\begin{figure}[H]
	\centering
	\includegraphics[width = 0.9\linewidth]{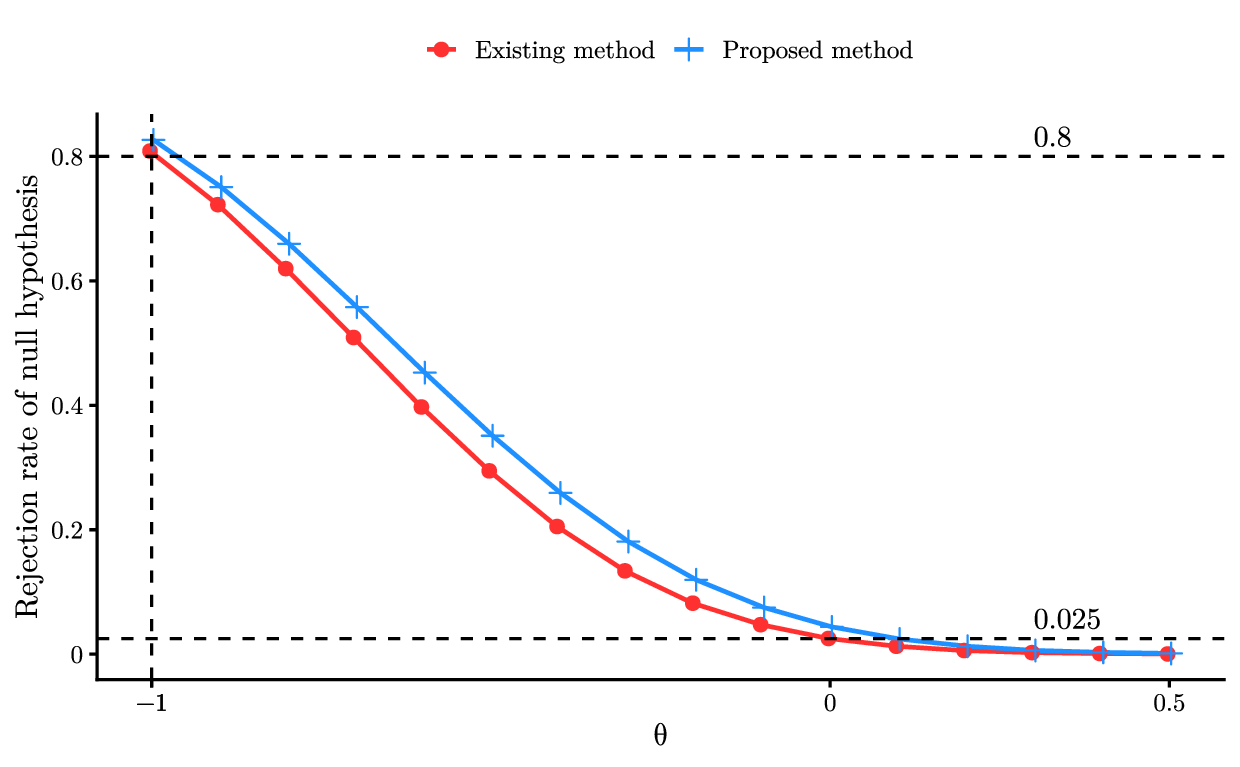}
	\caption{Rejection rates of the null hypothesis for $\bar y_\mathrm{E} = -1.0$ and $N_\mathrm{E} = 50$. The sample sizes $N_\mathrm{C}$ for the existing and proposed methods were 8 and 7, respectively.}
\end{figure}

\begin{figure}[H]
	\centering
	\includegraphics[width = 0.9\linewidth]{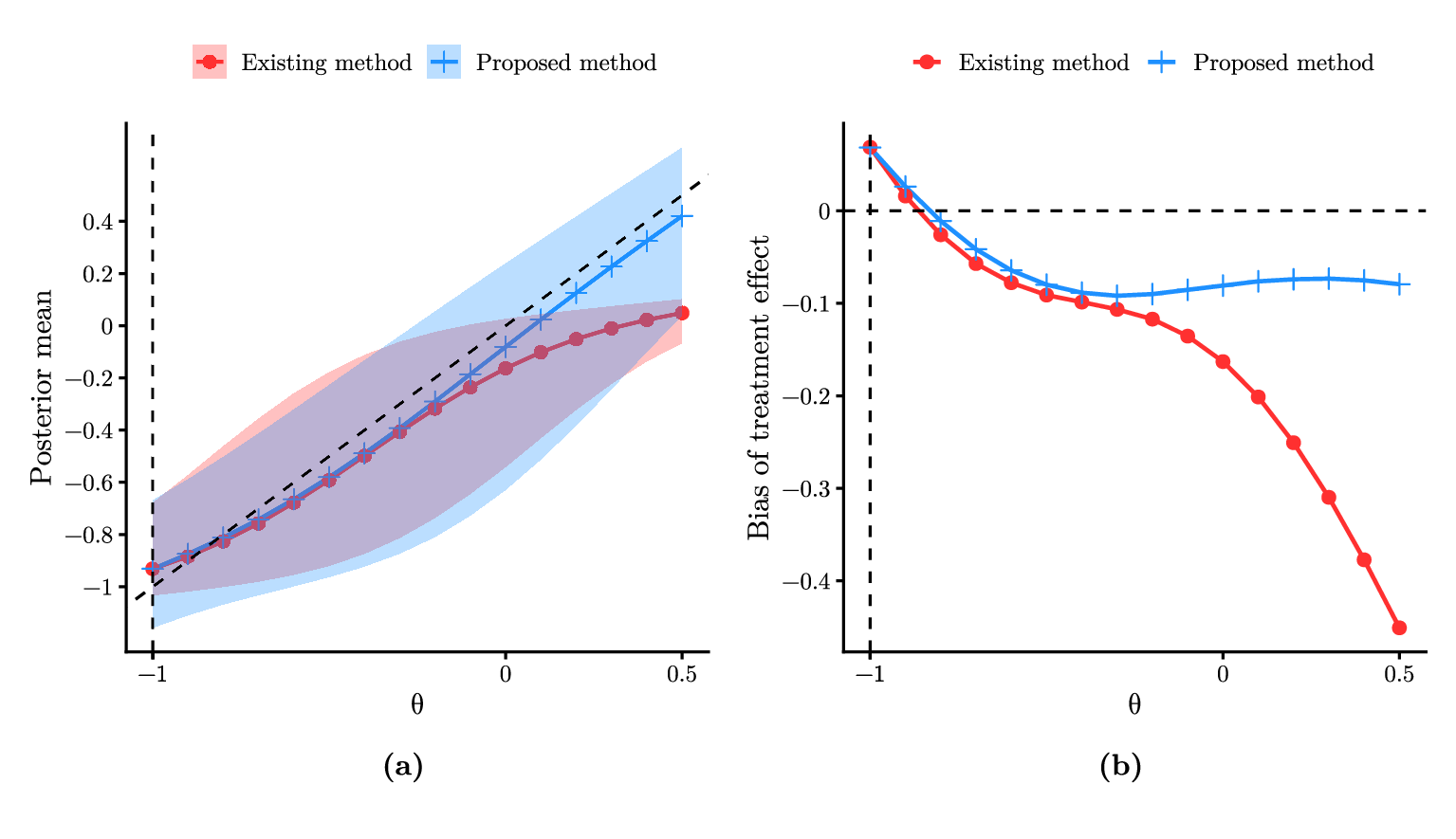}
	\caption{Posterior means, 95\% credible intervals (95\% CIs) and biases of the treatment effect estimates for $\bar y_\mathrm{E} = -1.0$ and $N_\mathrm{E} = 50$. The sample sizes $N_\mathrm{C}$ for the existing and proposed methods were 8 and 7, respectively.}
\end{figure}

\subsection{$\bar y_\mathrm{E} = -0.4$ and $N_\mathrm{E} = 200$}
\begin{figure}[H]
	\centering
	\includegraphics[width = 0.9\linewidth]{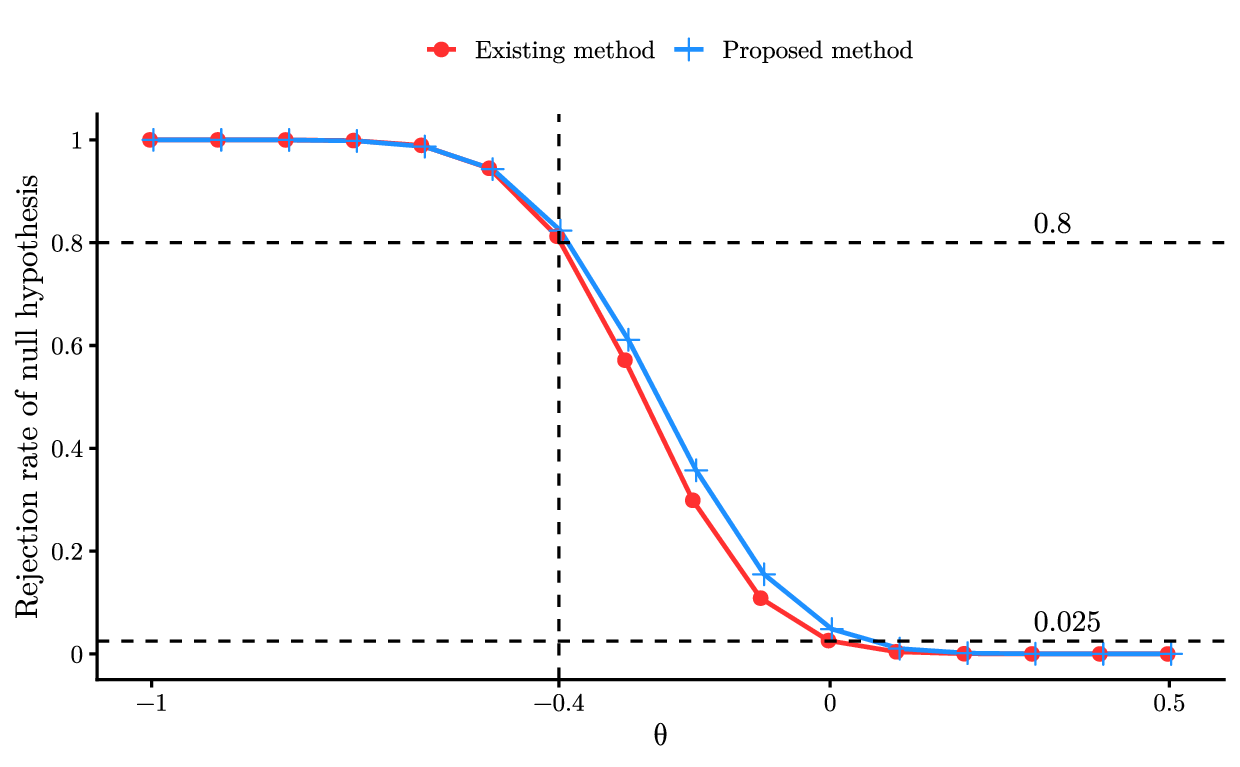}
	\caption{Rejection rates of the null hypothesis for $\bar y_\mathrm{E} = -0.4$ and $N_\mathrm{E} = 200$. The sample sizes $N_\mathrm{C}$ for the existing and proposed methods were 50 and 42, respectively.}

\end{figure}

\begin{figure}[H]
	\centering
	\includegraphics[width = 0.9\linewidth]{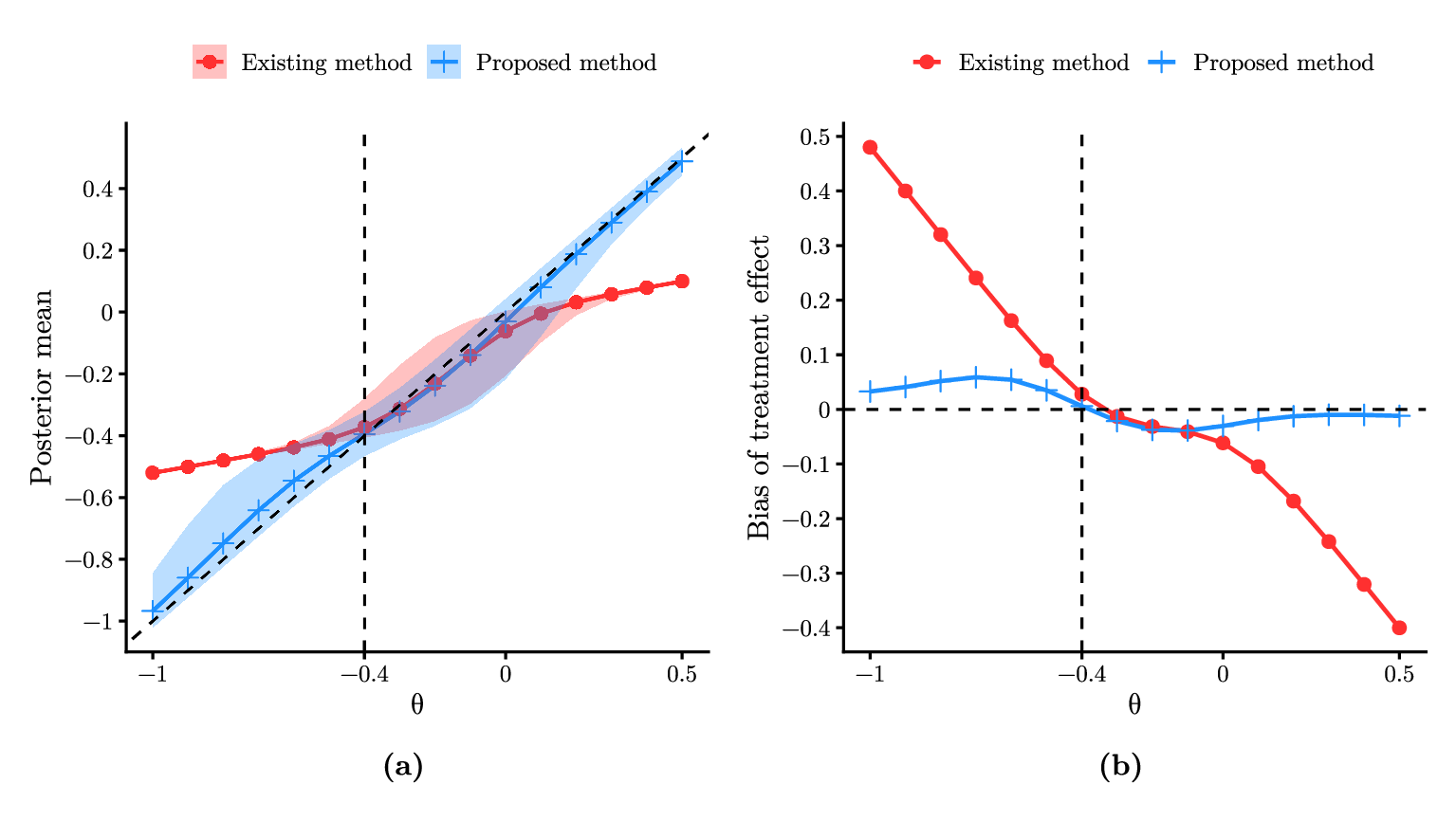}
	\caption{Posterior means, 95\% credible intervals (95\% CIs) and biases of the treatment effect estimates for $\bar y_\mathrm{E} = -0.4$ and $N_\mathrm{E} = 200$. The sample sizes $N_\mathrm{C}$ for the existing and proposed methods were 50 and 42, respectively.}
\end{figure}

\subsection{$\bar y_\mathrm{E} = -0.8$ and $N_\mathrm{E} = 200$}
\begin{figure}[H]
	\centering
	\includegraphics[width = 0.9\linewidth]{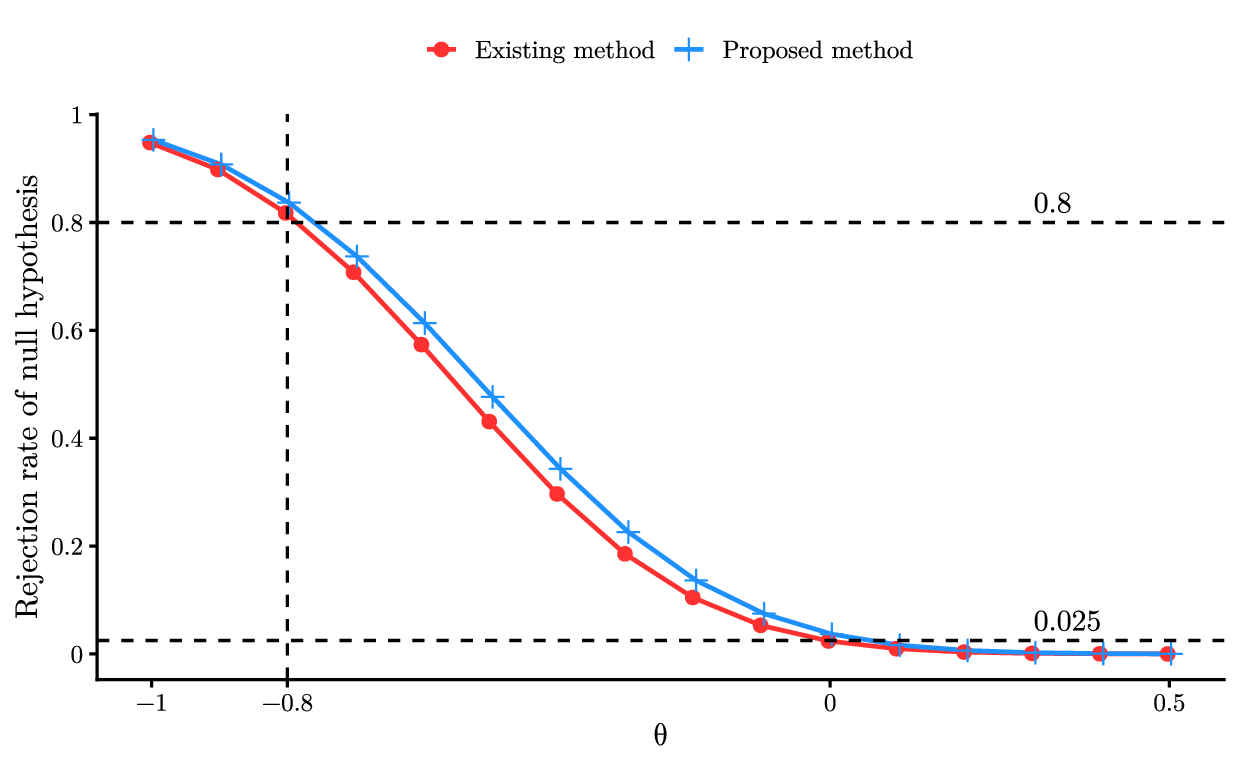}
	\caption{Rejection rates of the null hypothesis for $\bar y_\mathrm{E} = -0.8$ and $N_\mathrm{E} = 200$. The sample sizes $N_\mathrm{C}$ for the existing and proposed methods were 13 and 12, respectively.}
\end{figure}

\begin{figure}[H]
	\centering
	\includegraphics[width = 0.9\linewidth]{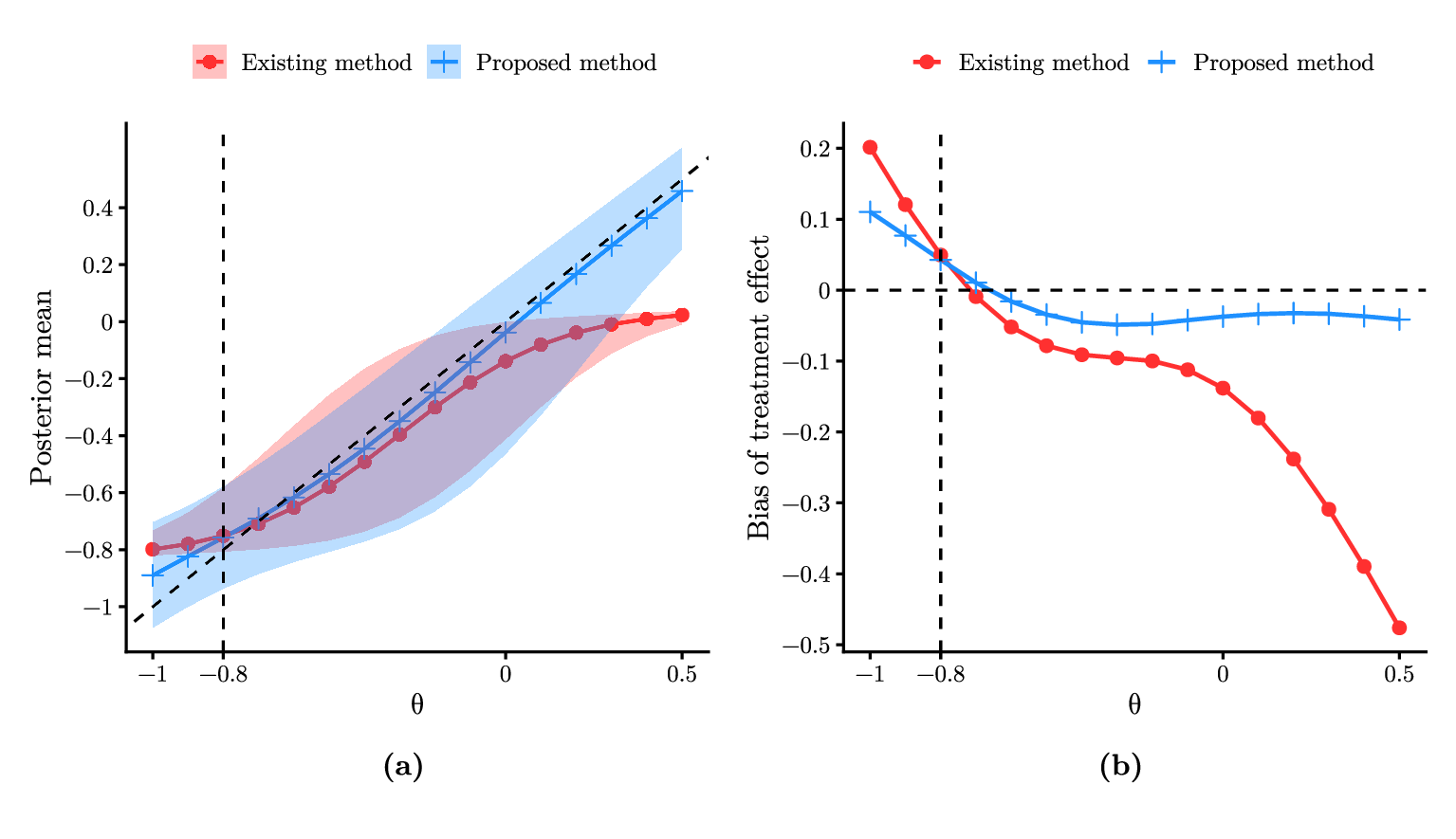}
	\caption{Posterior means, 95\% credible intervals (95\% CIs) and biases of the treatment effect estimates for $\bar y_\mathrm{E} = -0.8$ and $N_\mathrm{E} = 200$. The sample sizes $N_\mathrm{C}$ for the existing and proposed methods were 13 and 12, respectively.}
\end{figure}

\subsection{$\bar y_\mathrm{E} = -1.0$ and $N_\mathrm{E} = 200$}
\begin{figure}[H]
	\centering
	\includegraphics[width = 0.9\linewidth]{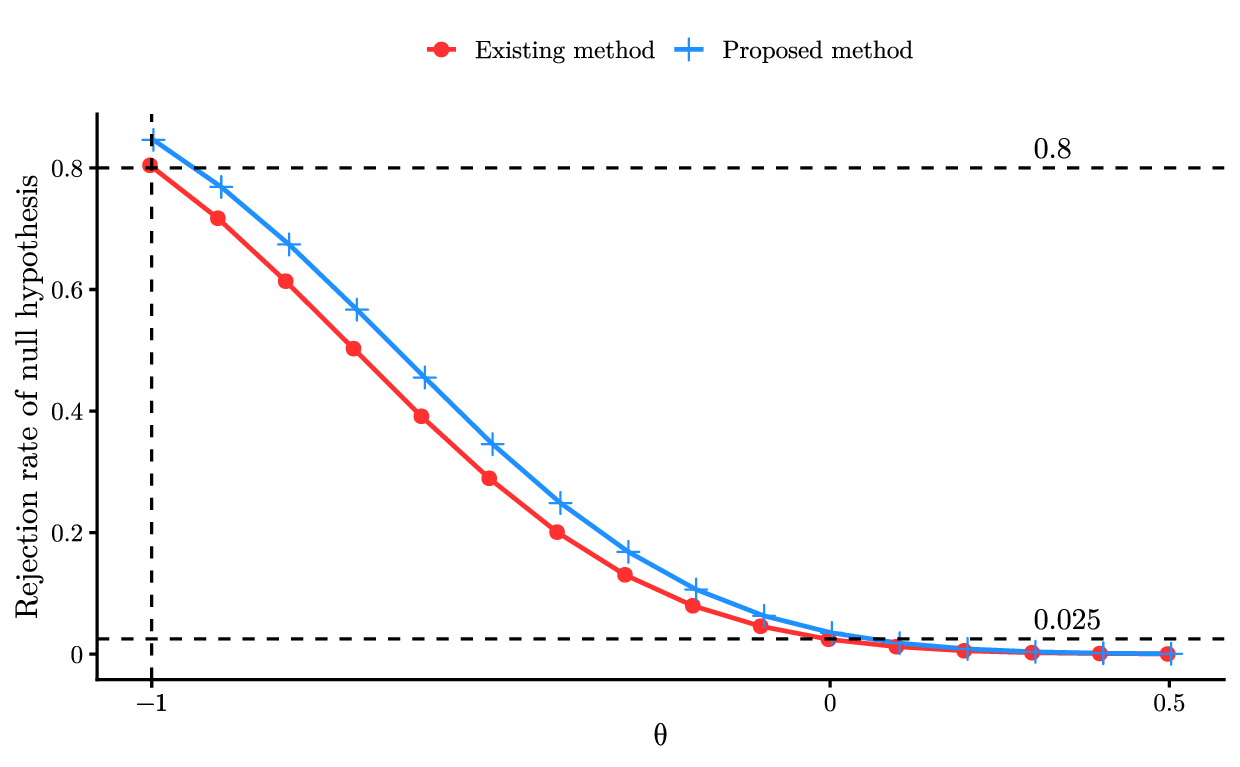}
	\caption{Rejection rates of the null hypothesis for $\bar y_\mathrm{E} = -1.0$ and $N_\mathrm{E} = 200$. The sample sizes $N_\mathrm{C}$ for the existing and proposed methods were 8.}
\end{figure}

\begin{figure}[H]
	\centering
	\includegraphics[width = 0.9\linewidth]{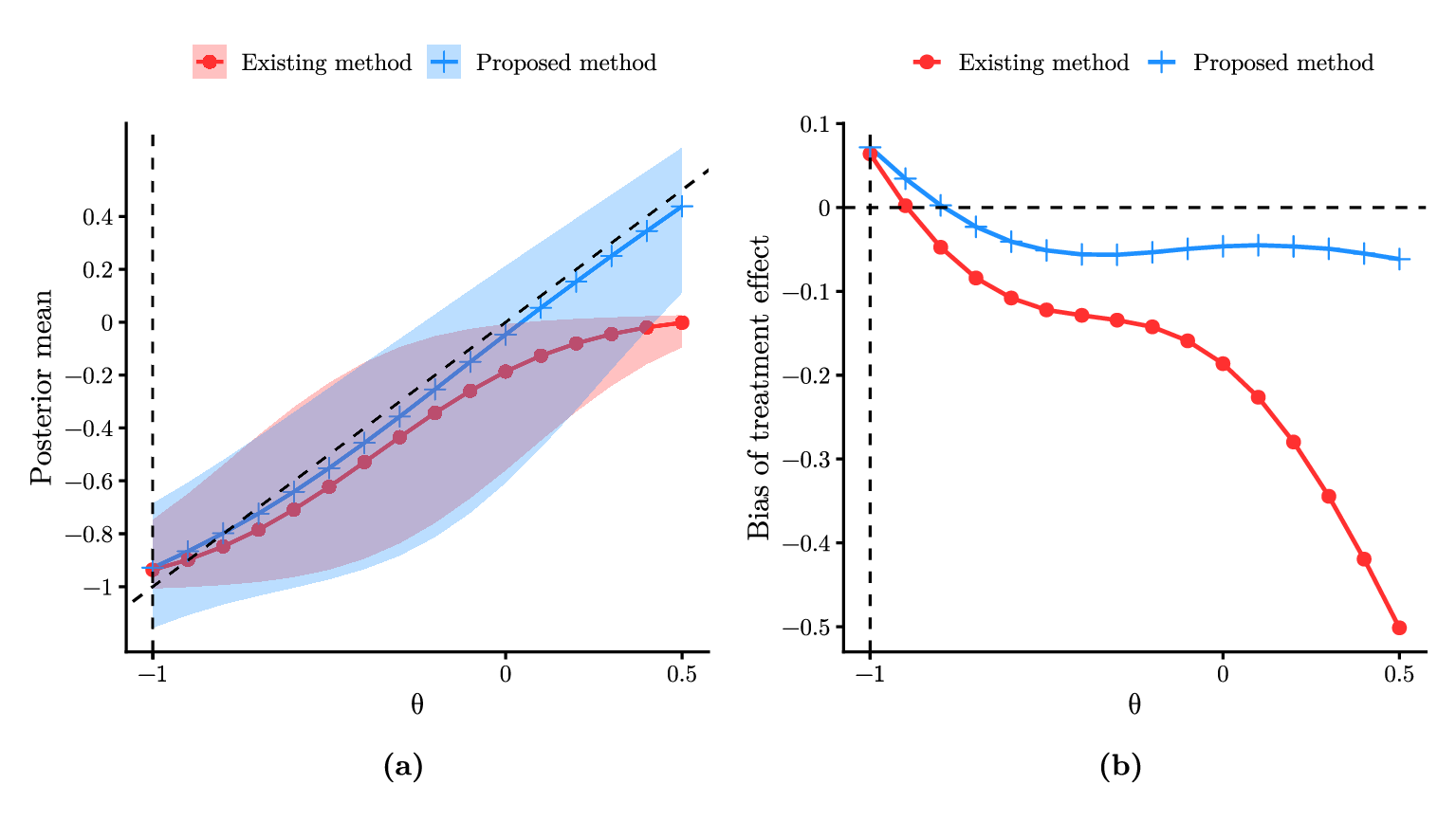}
	\caption{Posterior means, 95\% credible intervals (95\% CIs) and biases of the treatment effect estimates for $\bar y_\mathrm{E} = -1.0$ and $N_\mathrm{E} = 200$. The sample sizes $N_\mathrm{C}$ for the existing and proposed methods were 8, respectively.}
\end{figure}

\end{document}